\documentclass[twocolumn]{aastex631}
\usepackage{physics}

\def\gsim{\;\rlap{\lower 2.5pt
 \hbox{$\sim$}}\raise 1.5pt\hbox{$>$}\;}
\def\lsim{\;\rlap{\lower 2.5pt
   \hbox{$\sim$}}\raise 1.5pt\hbox{$<$}\;}

\begin{document}

\title{Tales of Tension: Magnetized Infalling Clouds and Cold Streams in the CGM}

\author[0009-0001-1399-2622]{Ish Kaul}
\affiliation{Department of Physics, University of California, Santa Barbara, CA 93106, USA}

\author[0000-0003-4805-6807]{Brent Tan}
\affiliation{Center for Computational Astrophysics, Flatiron Institute, New York, NY 10010, USA}

\author[0000-0002-1013-4657]{S. Peng Oh}
\affiliation{Department of Physics, University of California, Santa Barbara, CA 93106, USA}

\author[0000-0001-8057-5880]{Nir Mandelker}
\affiliation{Centre for Astrophysics and Planetary Science, Racah Institute of Physics, The Hebrew University, Jerusalem 91904, Israel}

\begin{abstract}
The observed star formation and wind outflow rates in galaxies suggest cold gas must be continually replenished via infalling clouds or streams. 
Previous studies have highlighted the importance of cooling-induced condensation on such gas, which enables survival, mass growth, and a drag force which typically exceeds hydrodynamic drag. However, the combined effects of magnetic fields, cooling, and infall remains unexplored. We conduct 3D magnetohydrodynamic (MHD) simulations of radiatively cooling infalling clouds and streams in uniform and stratified backgrounds. For infalling clouds, magnetic fields aligned with gravity do not impact cloud growth or dynamics significantly, although we see enhanced survival for stronger fields. By contrast, even weak transverse magnetic fields significantly slow cloud infall via magnetic drag, due to the development of strong draped fields which develop at peak infall velocity, before the cloud decelerates. Besides enhancing survival, long, slow infall {\it increases} total cloud mass growth compared to the hydrodynamic case, even if reduced turbulent mixing lowers the rate of mass growth. Streams often result in qualitatively different behavior. Mass growth and hence accretion drag are generally much lower in hydrodynamic streams. Unlike in clouds, aligned magnetic fields suppress mixing and thus both mass growth or loss. Transverse fields do apply magnetic drag and allow streams to grow, when the streams have a well-defined 'head' pushing through the surrounding medium. Overall, regardless of the efficacy of drag forces, streams are surprisingly robust in realistic potentials, as the destruction time when falling supersonically exceeds the infall time. We develop analytic models which reproduce cloud/stream trajectories.   
\end{abstract}



\section{Introduction} \label{sec:intro}
A major challenge in the theory of galaxy formation and evolution is the explanation of observed star formation rates, which are too high given the available gas reservoirs (assuming similar star formation rates across the age of the galaxy) and hence require continual replenishment in order to be sustained \citep{erb08,putman09}. Observations such as those from the Cosmic Origins Spectrograph (COS) installed on the Hubble Space Telescope (HST) have shed light on the importance of the hot gaseous halo surrounding galaxies, known as the circumgalactic medium (CGM), as a strong candidate for the source of this galactic nourishment. These observations reveal that the CGM is multiphase in nature, consisting of both hot and low density gas, as well as cold and high density gas, all of which are subject to complicated dynamics governed by turbulence and cooling/heating processes. The CGM thus plays a vital role in galaxy formation and evolution, being the medium through which various gas flows and physical mechanisms occur, thereby exchanging and recycling mass and energy between the galactic disk and itself \citep{Tumlinson2017,CGMReviewPeng}. 

Within this framework, the CGM is thought to be able to fuel new star formation within galaxies via two primary channels: gas inflows from cold clouds in the halo \citep{putman12} and cold streams \citep{keres05,DekelBrinboim2006,dekel09,Voort2012,mandelker20-blobs,Aung2024} from the cosmic web.
These two processes have thus been the subject of recent interest on the simulation front.
For the former, cold clouds are theorized to be ejected from the galaxy through large multiphase galactic winds and outflows driven by supernova feedback \citep{gronke18,FieldingBryanWinds,TanFielding2023}. 
Other possible sources of cold clouds include stripping from satellite galaxies \citep[e.g.,][]{roy24} and direct precipitation models \citep[e.g.,][]{voit19}. Regardless of origin, these cold clouds can then be accreted onto the galactic disk if they can survive the journey \citep{tan23-gravity}.

The other method of feeding galaxies with cold gas and sustaining star formation is via stream accretion. First seen in cosmological simulations, these narrow cold streams of high density cold gas follow dark matter filaments of the cosmic web, penetrate hot gaseous halos, and flow towards the central galaxies \citep{keres05,DekelBrinboim2006,dekel09,Voort2012}. Such cold streams are expected to commonly occur at redshifts $z>2$ in massive halos with $M_{\mathrm{vir}}>10^{12}M_{\odot}$. Due to their high densities and short cooling times, these streams maintain their structure with transonic inflow velocities with respect to the hot halo, potentially enabling them to reach the disk. While some cosmological studies \citep[e.g.,][]{GoerdtCeverino2015} have studied properties of the cold streams, such as their infall velocity and mass accretion rate, they remain limited by poor resolution, and hence such results remain sensitive to differences in numerical methods \citep{Nelson2013}. While all grid codes observe the survival of streams, there is little consensus on their properties or how quickly they might get disrupted.

The inflow mechanisms outlined above are not just features of simulation work. Much of the recent interest in this area has strongly been motivated by the latest observations of the CGM.
The cold phase of the CGM has been observed to have a plethora of $T\sim10^4$ K gas through emission and absorption lines (\cite{Veilleux2005}). There are many local observations related to high velocity clouds, whose existence and properties are well-established \citep{putman09, putman12}. 
However, observations of cold streams, from various emission lines of the CGM \citep{Turner2017,Zabl2020} and tomography of galactic winds \citep{Fu2021}, remain few in number due to their smaller covering fractions over the full CGM halo, so it is worth summarizing some observational motivation. There is strong direct evidence of cold filaments in quasars (\cite{Hennawi2015}) and those penetrating the galaxy at $z=2.9$ using Ly$\alpha$ narrow-band emissions \citep{Daddi2021}. Several other observations are consistent with filamentary structures in absorption and emission (\cite{Fumagalli2017}; \cite{Prochaska2014}; \cite{Umehata2019}; \cite{Bouche2016}; \cite{Cantalupo2014}; \cite{Martin2014}; \cite{Martin2014b}; \cite{Martin2019}; \cite{Matsuda2011}). More recently, a cold filamentary structure (\cite{EmontsColdStream}) has been observed at $z=3.8$ using the CI line, which traces atomic and molecular hydrogen. 

A running theme in recent works has been explaining the existence and survival of such cold gas structures as observed in the CGM, since they are thought to be particularly susceptible to being destroyed by instabilities \citep{klein94, zhang17}.
Modern refinements of this picture have centered around understanding the interaction of turbulent mixing and radiative cooling at the interfaces of the different phases \citep{ji18,fielding20,tan21}.
When this process is efficient, cloud-wind simulations have shown that entrainment of hot gas onto to the cold cloud is able to offset cloud destruction and can even lead to cold mass growth \citep{gronke18,gronke20-cloud}, if the cooling time is shorter than the cloud destruction time. Infalling cold clouds can also survive and grow, though they have a different survival criterion \citep{tan23-gravity}.
In a similar vein, previous work on cold streams have focused on using idealized wind tunnel setups (\cite{nirkhi2019}) coupled with radiative cooling (\cite{mandelker20}) or self-gravity (\cite{Aung2019}). 
the stream could grow in mass and survive. More recently, \cite{Aung2024} simulated these streams falling through an idealized halo gravitational potential coupled with radiative cooling, and found that these streams can penetrate the halo down to 0.1$R_{\mathrm{vir}}$. They also found radiation luminosities consistent with observations of Ly-$\alpha$ blobs in the CGM of galaxies at $z=2-4$.   

Despite these advances, one major source of uncertainty is the role of magnetic fields in this processes. 
While field strengths in the CGM are uncertain, plausible values range from $\beta = P_ {\rm thermal}/P_{\rm B} \sim 1-100$, with large scatter (e.g., \citealt{VandeVoort2021}). However, their role in determining the evolution and shaping the morphology of cold gas structures requires more work. 

How do magnetic fields affect infalling cold gas? 
As cold gas falls through the magnetized CGM, it sweeps up ambient traverse magnetic fields, which subsequently drape around the cold gas. This has several important effects on the dynamics and morphology of cold gas. From wind tunnel simulations it has been shown that the magnetic drag force acts to slow down the acceleration of the cloud by the wind by a factor of $\sim (1+\frac{2}{\beta_{\mathrm{wind}}\mathcal{M}^2})^{-1}$ (\cite{mccourt15}, \cite{dursi08}) with respect to the purely hydrodynamic case. In addition, in mixing layer simulations, magnetic tension acts to suppress mixing and cooling along the cold-hot interface \cite{ji19,zhao23}. This should have the natural implication that the cold mass growth rates from radiative cooling should be suppressed as well. However, \cite{gronke20-cloud} found that the growth rates are only marginally suppressed. More recent wind tunnel simulations (\cite{hildalgo23}) of cold clouds in hot winds have also found a dependence of the survival criteria of the cloud with $\beta$ as $t_{\rm{cool,mix}}/t_{\mathrm{cc}} \leq 1+200\beta^{-0.7}$. These imply that until $\beta$ is low enough to be order unity, the magnetic field only has marginal impacts on the overall dynamics of the cloud. This has remained largely unexplored in the context of infalling clouds and streams. 

Many of these simulations make idealizations it would be good to relax. For example, in \cite{mandelker20}, \cite{BerlokColdStream} and \cite{ColdStreamBfield}, the stream is initialized either as a cylinder \citep{mandelker20,BerlokColdStream} or 2D sheet \cite{ColdStreamBfield}) of cold dense gas with an initial velocity on the order of the hot gas sound speed moving through a hot uniform background with periodic boundary conditions in the flow direction. However, as streams pass through the CGM, they would face a gravitational potential forcing them to accelerate towards the galaxy. As shown by \cite{tan23-gravity} for clouds, this can radically change the survival criteria of streams. \cite{mandelker20} considered cooling in isolation with the stream and \cite{BerlokColdStream} considered magnetic fields in isolation with the stream, without cooling. While \cite{ColdStreamBfield} considered both of these effects along with thermal conduction, their results were limited to the 2D case and a very weak field of $10^{-3}\mu$G, which corresponds to a plasma beta $\beta = 10^5$. Similarly, previous studies on cold clouds have looked at magnetic fields only in wind tunnel scenarios \citep{gronke20-cloud,Li2020,Cottle2020}. \citet{tan23-gravity}) examined infalling cloud survival and growth in a stratified background, but only in hydrodynamics. \citet{gronnow22} did perform MHD simulations of cloud infall under gravity, but only ran the simulation for a short time, before many of the growth and drag effects we are interested in become important. They also did not examine survival criteria. 

In this work, we 
perform the first study of the influence of magnetic fields on the survival and growth of infalling, radiatively cooling clouds and streams.
We run a suite of 3D MHD simulations with radiative cooling for the two broad cases of the spherical cold cloud and a cylindrical cold stream falling under gravity in constant and stratified hot gas backgrounds. Many of these results differ significantly from wind tunnel simulations. 

The rest of the paper is structured as follows: we start by reviewing various timescales and analytic considerations for both cold clouds and cold streams in Section \ref{sec:theory}. Section \ref{sec:numerical_methods} provides a description of our numerical methods. In Section \ref{sec:test_theory}, we test the hydrodynamic theory with a more realistic setup. Sections \ref{sec:results_clouds} and \ref{sec:results_streams} then present the results of our cloud and stream simulations respectively. In Section \ref{sec:discussion}, we discuss the implications of our results within a wider astrophysical context. Finally, in Section \ref{sec:conclusions}, we conclude and provide directions for future research.

\section{Analytic Considerations} \label{sec:theory}
Before presenting our results, we first review the relevant body of analytical work for both the cold clouds and the cold streams. While most of this has been developed in the hydrodynamic limit (no magnetic fields), we will show that much of it remains applicable. At the end of the section, we discuss the impact of magnetic fields.

\subsection{Infalling Cold Clouds}
We begin by defining all of the relevant timescales for cold clouds. First, we have the cloud crushing time
\begin{equation}
    t_{\rm{cc}} = \sqrt{\delta}\frac{R}{v}
    \label{eq:tcc}
\end{equation}
where $R$ is the radius of the cloud, $v$ is the instantaneous velocity of the cloud with respect to the background medium, and $\delta$ is the density contrast between the cold cloud and the hot background. This defines the typical timescale for the destruction of the cloud due to internal shocks and hydrodynamic instabilities such as the Kelvin Helmholtz Instability (KHI) \citep{klein94}. In wind tunnel simulations, this velocity is usually defined to be the initial velocity of the wind, where the velocity difference is maximal. However, in a setup with a cloud falling under external gravity in a stationary background, we define this to be the actual instantaneous velocity of the cloud instead, since its velocity increases over time. 

\cite{gronke18} found that for wind tunnel simulations, the cloud is able to survive if
\begin{equation}
    t_{\rm{cool,mix}} < t_{\rm{cc}}
\end{equation}
where $t_{\rm{cool, mix}}$ is the cooling time of the mixed gas which has temperature $T_{\rm{mix}}\sim (T_{\rm{c}}T_{{\rm{h}}})^{1/2}$ and density $n_{mix}\sim (n_{\rm c}n_{\rm h})^{1/2}$, the geometric mean of the hot and cold gas temperatures and densities \citep{Begelman1990}. However, \cite{tan23-gravity} found that for the case of clouds falling under gravity, a more stringent criteria is required. Instead of the cooling time, the relevant time is the growth timescale
\begin{equation}
\label{eqn:cloud_tgrow}
    t_{\rm{grow}}\sim \frac{m}{\dot{m}}\sim \frac{V_{\rm cloud}\rho_{\rm{c}}}{\rho_{\rm{h}}A_{\rm cloud}v_{\rm{mix}}}\sim\delta\frac{R}{v_{\rm{mix}}}
\end{equation}
where $V_{\rm cloud}$ and $A_{\rm cloud}$ are the volume and surface area of the cloud, and $v_{\rm{mix}}$ (quantified later) is the hot gas inflow velocity, which depends on a combination of mixing and cooling, $\rho_{\rm{c}}$, $\rho_{\rm{h}}$ are the mass densities of the cold and hot gas, and $\delta = \rho_{\rm c}/\rho_{\rm h}$ is the cloud overdensity.
The criterion for survival is then \citep{tan23-gravity}
\begin{equation}
\label{eqn:cloud_survival}
    t_{\rm{grow}}<f_St_{\rm{cc}}
\end{equation}
where $f_S$ is a dimensionless constant of order unity ($\sim 4$ for uniform backgrounds and $\sim 1$ for stratified backgrounds, calibrated from simulations). We derive the full expression for $t_{\rm{grow}}$ later below.

We now summarize the analytical equations in \cite{tan23-gravity} for infalling cold clouds. We can begin by writing the equations that govern the dynamics of infalling clouds which and describe the evolution of cloud momentum, mass, and displacement. For a cloud of mass $m$ falling with velocity $v$ through a medium of density $\rho_{\mathrm{hot}}$, these are
\begin{equation}
    \label{eqn:momentum}
    \frac{d(mv)}{dt} = \dot{m}v + m\dot{v} = mg - \frac{\rho_{\mathrm{hot}} C_0 v^2A_{\rm cross}}{2}
\end{equation}
\begin{equation}
    \frac{dm}{dt} = \frac{m}{t_{\mathrm{grow}}(t)}
\end{equation}
\begin{equation}
    \frac{dz}{dt} = v
\label{eqn:dzdt}
\end{equation}
where the dotted quantities represent time derivatives, $z$ is the total distance the cloud has fallen, $g$ is the gravitational acceleration, $C_0$ is a geometry dependent drag coefficient, $A_{\rm cross}$ is the cross-sectional area perpendicular to the direction of motion, and $t_{\rm grow}$ is defined as the growth time of the cold gas that varies with time, whose scalings we shall derive below (based on \cite{tan23-gravity}). The last term of Equation \ref{eqn:momentum} is the conventional drag term, which turns out to be small compared to $\dot{m}v$ when the cloud or stream is growing \citep{brent25}.
The striking outcome of these equations is that accretion of background hot gas onto the existing cloud through mixing and cooling acts to reduce the momentum of the cloud during infall, referred to as an accretion-braking mechanism. This implies that the velocity of the cloud can be much lower than that expected from just ballistic free-fall, and in fact may even reach a terminal velocity as we shall show below. As mentioned above, $t_{\mathrm{grow}}\equiv m/\dot{m}$ depends on the mixing velocity $v_{\mathrm{mix}}$, which is derived from results of small scale turbulent mixing layer simulations \citep{tan21}: 
\begin{equation}
\label{eqn:vmix_vturb}
\resizebox{1.04\hsize}{!}{$v_{\rm mix} \approx 
    9.5 {\mathrm{~km}\,\mathrm{s}^{-1}}
    \left(\frac{u^{\prime}}{c_{\rm s, cold}}\right)^{3/4} 
    \left(\frac{L_{\rm turb}}{100\,{\rm pc}}\right)^{1/4}
    \left(\frac{t_{\rm cool}}{0.03\,{\rm Myr}}\right)^{-1/4}$}
\end{equation}
where $t_{\mathrm{cool}}$ is the {\it minimum} cooling time in the temperature range from $T_{\rm{h}}$ to $T_{\rm{c}}$\footnote{For isobaric cooling, cooling peaks at close to $T_{\rm{c}}$.}, $u^{\prime}$ is the peak turbulent velocity in the mixing layer, $L_{\rm turb}$ is the relevant outer scale of the turbulence, and $c_{\rm s, cold}=15 \, \rm{km\, s^{-1}} $ is the cold gas sound speed. For our purposes $L_{\rm turb}$ is of order the cloud radius since these are the largest turbulent eddies mixing the cloud surface. Shearing turbulent mixing simulations find that \citep{tan21}
\begin{equation}
    u' \approx 50{\mathrm{~km}\,\mathrm{s}^{-1}}
    \mathcal{M}^{4/5}
    \left(\frac{c_{\rm s,hot}}{150 {\mathrm{~km}\,\mathrm{s}^{-1}}} \right)^{4/5}
    \left(\frac{t_{\rm cool}}{0.03\,{\rm Myr}}\right)^{-0.1}
    \label{eq:uprime} 
\end{equation}
for $\delta\gtrsim 100$ and $\mathcal{M}\equiv v_{\rm shear} / c_{\rm s,hot}$ (where $u'=50 \rm{km s^{-1}}$ is the value found for this choice of normalization). We will neglect the weak $t_{\rm cool}$ dependence here going forward. \citet{YangJi2023} found that $u'$ saturates beyond $\mathcal{M} = 1$, which is accounted for by setting $\mathcal{M} \rightarrow \min(1,\mathcal{M})$.
A time dependent weight pre-factor
\begin{equation} 
    w_{\rm kh}(t) = \min\left(1, \frac{t}{f_{\rm kh} t_{\rm cc}}\right),
    \label{eqn:w_kh}
\end{equation} 
is also multiplied ($u'w_{\rm kh}(t)$) to account for the initial onset of turbulence. Lastly, the cloud surface area is given by
\begin{equation}
\label{eqn:cloud_surface_area}
    A_{\rm cloud} \approx A_{\rm cloud,0} \left( \frac{m}{m_0} \frac{\rho_{\rm cloud,0}}{\rho_{\rm cloud}} \right)^{5/6}
\end{equation}
where $A_{\rm cloud,0}$, $\rho_{\rm cloud,0}$ and $m_0$ are the initial cloud surface area, density and mass respectively at the initial galactocentric radius. $A_{\rm cloud}$, $\rho_{\rm cloud}$ and $m$ are the same parameters when the cloud has reached some galactocentric radius $r$. The change here is due to the underlying pressure profile in the halo, with the cold and hot temperatures, and density contrast always assumed to be the same (isothermal conditions). Equation \ref{eqn:cloud_surface_area} is a numerical fit to the simulations; the fact that $A_{\rm cloud}\propto m^{\alpha}$ where $\alpha \approx 5/6 >2/3$ is due to the fractal nature of the cloud interface. Recall the expression for the growth time, rewritten to reflect all the quantities at stake
\begin{equation}
    t_{\rm grow,0} = \frac{m_0}{\rho_{h,0} A_{\rm cl,0} v_{\rm mix,0}}
\end{equation}

Combining all of the scalings for $v_{\rm mix}$ with the approximation $L \sim R$ and substituting for $u'$  gives an expression for the growth time
\begin{align}
    t_{\rm grow} = \frac{t_{\rm grow,0}}{w_{\rm kh}(t)}
                   \left( \frac{c_{\rm s,hot, 150}}{v} \right)^{3/5}
                   \left( \frac{t_{\rm cool}}{t_{\rm cool,0}} \right)^{1/4}
                   \left( \frac{m}{m_0} \frac{\rho_{\rm hot,0}}{\rho_{\rm hot}} \right)^{1-\alpha},
    \label{eq:t_grow_cloud}
\end{align}
where $c_{\rm s,hot, 150} = 150\,{\rm km\,s}^{-1}$ is the sound speed of gas at $10^6$~K and the initial growth time $t_{\rm grow,0}$ is given by
\begin{align}
\resizebox{1.04\hsize}{!}{$
    t_{\rm grow,0} \approx 35\,{\rm Myr }\,
                     \left( \frac{f_{\rm A}}{0.23} \right)
                     \left( \frac{\delta}{100} \right)
                     \left(\frac{R}{R_{100}}\right)^{3/4}
                     \left( \frac{t_{\rm cool,0}}{0.03\,{\rm Myr}} \right)^{1/4}.
$}
\end{align}
where $f_{\mathrm{A}}$ is a small constant factor set by calibrating simulations, and $R_{100}=100\ \mathrm{pc}$.
As the cloud falls through the medium, it starts out with ballistic infall, i.e. $v\sim gt$ because turbulence has not set in. However, once mixing and cooling becomes effective, the cloud's velocity saturates to $v_{T} \sim gt_{\mathrm{grow}}$ (which arises from balancing the two dominant terms in equation \ref{eqn:momentum}, $\dot{m} v \sim m g$). 
Interestingly, if we evaluate Equation \ref{eq:t_grow_cloud} at the terminal velocity of $v_T\sim gt_{\mathrm{grow}}$ and ignore the weak mass dependence, we find that $t_{\mathrm{grow}} \propto R^{15/32}$ which implies that the ratio $t_{\mathrm{grow}}/t_{\mathrm{cc}} \propto \sqrt{\delta}gt_{\rm{grow}}^2/R\propto R^{-1/16}$ is almost independent of the cloud radius. The survival criteria $t_{\mathrm{grow}}<4t_{\mathrm{cc}}$ can be thus written in a form expressing that the pressure (or density) be high enough that the cooling time is short, irrespective of cloud size. This yields \citep{tan23-gravity}
\begin{equation}
    P>3000k_B \, \mathrm{K \, cm^{-3}}\left(\frac{g}{10^{-8}\mathrm{cm \, s^{-2}}}\right)^{4/5}\left(\frac{\delta}{100}\right)^{12/5}
\end{equation}
In other words, with a stratified background, cold clouds only have to survive infall long enough to enter a sufficiently dense region where cooling times are short enough, where they can then grow in mass and reach the disk.

In this paper, we examine two magnetic geometries, with B-fields either aligned or transverse to the infall direction. As we shall see, aligned fields have little effect; the MHD simulations largely mirror hydrodynamic results. Aligned fields do not exert strong MHD forces, or significantly inhibit mixing. However, when the cloud infall includes magnetic fields transverse to the infall, the cloud drapes magnetic fields around it which get compressionally amplified. There is an additional drag term due to magnetic tension in this draping region such that the total drag in this case becomes \citep{dursi08, mccourt15}
\begin{equation}
    F_{\rm drag} \sim \rho_{\rm hot} v^2 R^2 \left[ 1+ \left(\frac{v_{\rm A}}{v}\right)^2 \right] 
    \label{eqn:Fdrag}
\end{equation}
where $v_A$ is the Alfven velocity in the draped region. This modifies Equation \ref{eqn:momentum} to include the term:
\begin{equation}
\label{eqn:momentum_with_mag}
    \frac{d (mv)}{dt} = \dot{m}v+m\dot{v} = mg - F_{\rm drag}
\end{equation}
where $F_{\rm drag}$ now includes the magnetic and hydrodynamic drag. For sufficiently low Alfven Mach number $\mathcal{M_A}\equiv v/v_A$, this would become a dominant term in the momentum equation, to the point where the steady state equation becomes
\begin{equation}
    mg = \rho_{\rm hot} v^2R^2\left[1+\left(\frac{v_A}{v}\right)^2\right]
\end{equation}
which means that instead of having $v_{T} = gt_{\rm grow}$, we get a new terminal velocity that depends on the magnetic drag term
\begin{equation}
\label{eqn:v_T_mag}
    v_{\rm T,mag} = \left(\frac{mg}{\rho_{\rm hot} R^2}-v_A^2\right)^{1/2}\sim(\delta Rg - v_A^2)^{1/2}
\end{equation}
As we shall see later, this turns out to be an excellent estimate for clouds once they get to low $\beta_{\rm drape}$.

\subsection{Infalling Cold Streams}
Observations and simulations suggest that cold streams are most typically present in massive halos with masses $M_h>10^{12}M_{\odot}$ and at redshifts $z\sim2$ to $z\sim4$. Under a standard spherical collapse model for an EdS universe with $\Omega_{\rm{m}}=1$ (which is true at high redshifts), one can derive analytic expressions for the galactic halos through which streams pass, relating their virial radius and velocity to redshift \citep{dekel2013}:
\begin{equation}
    R_{\rm{vir}} = 100 \rm{kpc} \left(\frac{M_h}{10^{12}M_{\odot}}\right)^{1/3}\left(\frac{1+z}{3}\right)^{-1}
\end{equation}

\begin{equation}
    V_{\rm{vir}} = 200\rm{km} \, \rm{s}^{-1} \left(\frac{M_h}{10^{12}M_{\odot}}\right)^{1/3}\left(\frac{1+z}{3}\right)^{-1/2}
\end{equation}
where $M_h$ is the virial mass of the halo. Typically, these streams have number densities $n_s\sim 10^{-1} - 10^{-3} \rm{cm}^{-3}$ and low metallicities $Z_s \sim 10^{-2} - 10^{-1} Z_{\odot}$ in cosmological simulations \citep{dekel2013,Goerdt2010,mandelker20-blobs,ColdStreamBfield} or in analytic derivations \citep{mandelker20-blobs}. In addition the density contrast is constrained to $\delta\sim30-300$, Mach number $\mathcal{M}\sim0.75 - 2.25$ (ranging from subsonic to supersonic), and radius $R_s\sim 3-50 $ kpc.

The analogue of the cloud crushing time $t_{\rm{cc}}$ for streams is known as the shearing time $t_{\rm{sh}}$. This represents the timescale on which KHI (due to shear along the stream boundaries) disrupts the stream, i.e. the time taken for the shear layer to grow to the size of stream radius \citep{Aung2019}. Using idealized simulations, \cite{nirkhi2019} and \cite{mandelker20} found an expression for this shearing time for the case of hydrodynamic cold streams during the early stream evolution without cooling:
\begin{equation}
    \label{eqn:tsh}
    t_{\rm{sh}} = \frac{R_{s,0}}{\alpha v_s}
\end{equation}
where $R_{s,0}$ is the  radius of the stream, $v_s$ is the velocity of the stream, and 
\begin{equation}
    \alpha = 0.21 \times (0.8\exp{( -3\mathcal{M}_{\rm{tot}}^2)} + 0.2)    
\end{equation}
is the growth rate of KHI in the shear layer found empirically by \cite{dimotakis} for the 2D growth of KHI. The total Mach number in the expression is given by $\mathcal{M}_{\rm{tot}} = v_s/(c_{\rm{s,cold}} + c_{\rm{s,hot}})$. 
This growth time has been checked in 3D stream simulations for astrophysically relevant parameters \citep{mandelker19}. 
\cite{mandelker20-blobs} extended the theory of stream evolution and entrainment from wind tunnel simulations to the case of a stream falling through a realistic halo potential. They predicted that the stream gets squeezed while falling through the potential, leading to lower surface area and higher density as it reaches closer to the galactic center. This also causes an increase in the total emitted luminosity to $\mathcal{L} \sim 10^{42} \ \rm{erg \ s^{-1}}$. This could potentially help explain the existence of Ly$\alpha$ blobs which have similar luminosities. This theoretical prediction was confirmed by 3D hydro simulations with radiative cooling of the stream falling through an NFW potential through a $10^{12} M_{\odot}$ halo in \cite{Aung2024}. There, they found stream profiles for temperature, density, velocity, mass entrainment rate and emitted luminosity that match the earlier predictions.
  
There exist two main types of instabilities in moving 3D cylindrical streams: surface modes, which occur as perturbations on the surface of the stream, and body modes, which occur as large scale sinusoidal motions across the entire body of the stream \citep{ mandelker16,padnos18,nirkhi2019}. It is worth highlighting that the expression for the shear time in Equation \ref{eqn:tsh}, which is based on the growth timescale of surface modes, would likely not hold for highly supersonic velocities. This is because for high Mach numbers, Equation \ref{eqn:tsh} would give  $t_{\rm{sh}}\propto 1/v$ since $\alpha$ tends to a constant. This becomes unphysical since the stream cannot disrupt in less than the sound crossing time of the stream $t_{\mathrm{sc}}$, which is defined as
\begin{equation}
    t_{\rm{sc}} = \frac{2R_s}{c_{\rm{s,cold}}}
\end{equation}
and represents the typical timescale for information to travel across the entire stream. This will also serve as a characteristic timescale in our numerical simulations.

\cite{mandelker20} theorize that, similar to the cloud case, survival is dictated by a competition between a growth timescale $t_{\rm{cool,mix}}$ and a destruction timescale $t_{\rm{sh}}$. 
Equating $t_{\rm{cool,mix}}$ and $t_{\rm{sh}}$ gives\footnote{Since $\alpha$ is a function of stream velocity, albeit a very weak one, $\alpha$ and $v_s$ are not completely independent of each other}:
\begin{equation}
    R_{\rm{s,crit}} = 20\rm{pc}\left(\frac{\alpha}{0.05}\right)\mathcal{M_{\rm hot}}\left(\frac{t_{cool,mix}}{1.9 \rm{Myr}}\right)\left(\frac{\delta}{100}\right)^{1/2}
\end{equation}
Streams with radii greater than $R_{\rm{s,crit}}$ will survive and grow in mass since that implies $t_{\rm{cool,mix}} < t_{\rm{sh}}$. We note that the critical radius here is smaller than the one in \cite{mandelker20} by an order of magnitude. This discrepancy is due to the difference in metallicities adopted (here $Z=Z_{\odot}$ compared to $Z=0.1Z_{\odot}$ for background and $Z=0.03Z_{\odot}$ for the stream in \cite{mandelker20}) along with the inclusion of background UV heating, which affect the cooling rates (and thus $t_{\rm{cool,mix}}$) by approximately that factor.

In order to extend equations \ref{eqn:momentum}-\ref{eqn:dzdt} to a cold stream falling under external gravity, we follow an equivalent derivation to the one outlined above for infalling clouds.
Instead of tracking cloud properties as they evolve over time, we derive spatial profiles of an inflowing stream which has a constant source of cold material.
We can thus write down the conservation equations for mass and momentum: 
\begin{eqnarray}
\frac{\partial \rho}{\partial t} + \nabla \cdot \left(\rho {\mathbf v} \right) &=& \left[ \frac{d \, \rho}{d\, t}\right] \approx \frac{\rho}{t_{\rm grow}}
\label{eq:stream_analytic1}
\\
\frac{\partial \rho \mathbf{v}}{\partial t} + \nabla \cdot \left(\rho \mathbf{vv} \right) &=&
\mathbf{f}
\approx \rho \mathbf{g}
\label{eq:stream_analytic2}
\end{eqnarray}
where $[d\rho/dt], \mathbf{f}$ are the mass and momentum source terms respectively. These represent growth in stream mass (due to condensation of hot halo gas) and forces $\mathbf{f}$ acting on the stream. We adopt approximations for these source terms which we discuss below. For now, if we adopt these approximations, assume that all variations occur in the z direction, and note that the divergence operator for the stream takes the form\footnote{Similar to the form for a magnetic flux tube, e.g. \citet{breitschwerdt91}.} $\nabla \cdot = A^{-1} d/dz (A \cdot )$, where $A$ is the cross-sectional area of the stream. We obtain in steady state: 
\begin{equation}
\label{eqn:stream_eq1}
    \frac{d}{dz}(\rho Av) = \frac{\rho A}{t_{\rm{grow}}}
\end{equation}
\begin{equation}
\label{eqn:stream_eq2}
    \frac{d}{dz}(\rho Av^2) = \rho Ag
\end{equation}
 The stream density $\rho(z)$ can be determined by the background hot gas pressure profile $P(z)$, if the stream is at $T_{\rm c} \sim 10^4$K and is in pressure balance with the hot gas: $\rho \propto P_{\rm hot}/T_c$. However, this is no longer true once magnetic pressure support becomes important. 

The stream growth timescale\footnote{Note that \cite{Aung2024} and \cite{mandelker20-blobs} use a similar quantity, the entrainment time $t_{\rm{ent}}$, except that the turbulent velocity in the mixing layer is assumed to be the cold gas sound speed, $u^{\prime} \rightarrow c_{\rm s,c}$. While this is the right order of magnitude, in detail the turbulent velocity depends on the shear (i.e., the infall velocity).} is: 
\begin{equation}
    t_{\rm{grow}}(z) \sim \frac{m}{\dot{m}}\sim \frac{\rho_{\rm{cold}}(\pi R_s^2 \dd z)}{\rho_{\rm{hot}}(2\pi R_s \dd z)v_{\rm{mix}}} \sim \delta\frac{R_s}{v_{\rm{mix}}}
\end{equation}
where $\delta$ is the density contrast of the cold stream with the hot background, and  $R_s$ is the stream radius. This is directly analogous to Equation \ref{eqn:cloud_tgrow} for clouds. Here we use the scalings as motivated by infalling clouds \citep{tan23-gravity} where the relevant time scale for determining survival and growth is $t_{\rm{grow}}$.
As with the infalling clouds, by using Equations \ref{eqn:vmix_vturb} and \ref{eq:uprime} for mixing velocity, we obtain a similar expression for $t_{\rm{grow}}$:
\begin{equation}
    \resizebox{1.04\hsize}{!}{$t_{\rm{grow}} = \frac{35\rm{Myr}}{w_{\rm{kh}}(z)}\left(\frac{f_{\rm{A}}}{0.17}\right)\left(\frac{c_{s,150}}{v}\right)^{3/5}\left(\frac{\delta}{100}\right)\left(\frac{R_s}{R_{100}}\right)^{3/4}\left(\frac{t_{\rm{cool}}}{0.03\rm{Myr}}\right)^{1/4}$}
    \label{eq:tgrow_stream}
\end{equation}
where $f_A$ is calibrated from injected stream simulations. One notable difference is that unlike the cloud which undergoes a change in geometry by growing a tail, the stream has no such evolution. Equation \ref{eqn:cloud_surface_area}, which expresses the growth in area with mass, is thus not relevant in this derivation. In the stream geometry, surface area is roughly constant throughout the simulation.
The other difference arising from the setup is the formulation of $w_{\rm{kh}}$ (equation \ref{eqn:w_kh}), which accounts for the onset of turbulence. Since $t_{\rm{sh}}$ is defined precisely as this timescale for streams, we use instead:
\begin{equation}
    w_{\rm{kh}} = \rm{min}\left(1, \frac{t}{f_{\rm{kh}}t_{\rm{sh}}}\right) = \rm{min}\left(1, \frac{\alpha z}{f_{\rm{kh}} R_s}\right).
\end{equation}
As for the forces acting on the stream, $\mathbf{f}$, we have only considered gravitational forces, and ignored thermal pressure gradients $\nabla P$ (which are negligible for cold streams compared to the gravitational force) and MHD forces. We shall later see that the latter can be important for some magnetic geometries.

 
\section{Numerical Methods} \label{sec:numerical_methods}
We perform 3D numerical hydrodynamic and magnetohydrodynamic (MHD) simulations using the publicly available Athena++ code (\citealt{Athena}). The simulations are performed on a regular Cartesian grid. We use the HLLC Riemann solver for hydrodynamic cases and switch to using a HLLD solver for the MHD cases. We also use the Piecewise Linear Method (PLM) applied to primitive variables for second order spatial reconstruction. We use the third-order accurate Runge-Kutta method for time integration.

We perform our simulations in elongated rectangular boxes where the $z$ direction is longer than the $x$ and $y$ directions to properly capture the tails of clouds and the dynamics of streams. 
Unless otherwise stated, we utilize $256\times256 \times 1024$ boxes of width 4.8 kpc (in $x$, $y$), and height 19.6 kpc (in $z$), corresponding to a resolution of $\sim 19$ pc. Our fiducial clouds and streams have radii $r \sim 300$ pc\footnote{while this is quite narrow for astrophysical streams \citep{mandelker20-blobs}, we use this radius for consistency with cloud simulations}. Thus, each cloud/stream is resolved by $\sim 32$ cells across its diameter (16 cells per radius). Previous work on both clouds \citep{gronke20-cloud,hildalgo23} and streams \citep{mandelker20} find that this is more than sufficient to establish converged mass evolution histories, at least in the hydrodynamic case. Later we will also show that the mass growth histories are converged at this resolution when magnetic fields are included.

{\it Uniform Background.} A uniform background in the presence of gravity is of course unrealistic, as there are no pressure gradients to oppose gravitational forces\footnote{In practice, we implement this by adding an artificial upward force $\rho_{\rm h} g$ at every point. This allows the hot gas to remain static, but has virtually no impact on colder gas, since $\rho_{\rm c} \gg \rho_{\rm h}$.}. However, this idealization allows us to understand the dynamics of infalling clouds and streams with no change in cooling strengths due to stratification of the background.
Here, the background is filled uniformly with a hot gas of $T_{\rm{hot}}=10^6$ K and number density $n = 10^{-4}\rm{cm}^{-3}$.  The cloud and stream are initialized with a cold gas of $T_{\rm{cold}}=10^4$ K with density $n = 10^{-2} \rm{cm}^{-3}$ and are hence in pressure balance with the background (even though ionization states at the two temperatures are slightly different, this is a small effect and we thus assume $\mu=1$ everywhere here). This gives a density contrast of $\delta = 100$ which we keep fixed for all simulations. 
We also randomly perturb the velocity profiles with velocity fluctuations $\sim 1\%$ of the hot gas sound speed to seed fluid instabilities. 
This setup has an initial sharp discontinuity in density and temperature at the interface of the cold gas (stream and cloud) and the hot background. We use outflow boundary conditions in the $x,y$ directions. For clouds, we adopt inflow and outflow conditions at the bottom and top of the z boundary, respectively.
We also perform frame tracking, where at each time-step, we perform a Galilean boost to remain in the frame of the cloud center of mass in order to properly capture the cometary tail \citep{mccourt15,gronke18,tan23-gravity}. For streams in a uniform background, we use periodic boundary conditions in z as in \cite{mandelker20}. We assume constant acceleration due to gravity with a value $g = 10^{-8}\rm{cm}/\rm{s}^2$ as provided by a fit of the Milky Way from \cite{BenjaminDanly}; see \citet{tan23-gravity} for further discussion.

{\it Stratified Background, Constant Gravity.} The stratified background case reflects a more realistic scenario of a hydrostatic isothermal gas where pressure gradients balance gravitational forces on the hot gas and prevent it from falling. We first consider a constant $g=10^{-8}\rm{cm}/\rm{s}^2$, as appropriate near the disk, and initialize the background to have a density profile: 
\begin{equation}
    n_{\rm{CGM}}(z) = n_{0}\exp{\left(-\frac{z}{H}\right)}
\end{equation}
where $n_0 = 10^{-4}\rm{cm}^{-3}$ and the isothermal scale height is $H = k_BT/\mu m_Hg = 2.7 \, \rm{kpc}$. In this framework, the cloud is dropped starting at $z=0$. Thus, the density increases and cooling time decreases as the cold material falls through the stratified background. Note that constant gravity is only appropriate close to the disk, perhaps for $\sim 2-3$ scale heights. We simulate many more scale heights than that, as a numerical experiment to tease out the underlying physics, similar to the approach in \citet{tan23-gravity}. This enables us to build an analytic understanding which we then validate with simulations using fully realistic background profiles (see below). 

For clouds, we continue to use a cloud-tracking algorithm, as in the uniform background case. In the stream case, where we are simulating the properties of a long (semi-infinite) cylinder, frame tracking is no longer appropriate. Instead, for the uniform background case, similar to stream wind tunnels \citep{mandelker20}, we use periodic boundary conditions at the top and and bottom. However, for the stratified case, periodic boundary conditions are no longer appropriate. Instead, as in \citet{Aung2024}, we continually inject cold gas stream material at the top of the box at $v_z=V_{\rm{vir}}=200\ \rm{km/s}$, and allow it to fall to the bottom of the box under gravity. At the top of the box, the hot background is at $T=10^6 K$ with $n=10^{-4}\rm{cm^{-3}}$, and the cold material is at $T=10^4$ K and initial density $n=10^{-2}\rm{cm}^{-3}$.
While we use inflow boundary conditions at the top, at the bottom we use hydrostatic boundary conditions to specify density and pressure in the ghost zones. 

{\it Stratified Background, Varying Gravity.} Finally, we consider cases where gravity $g(r)$, as well as hot halo gas density and temperature profiles $n(r), T(r)$ are more realistic. The cold gas streams/clouds do not fall for an unrealistic number of pressure scale heights, but instead follow trajectories consistent with falling from a given radius towards the halo center. We adapt the profiles $g(r), n(r), T(r)$ which would appear in spherical geometry to equivalent profiles $g(z), n(z), T(z)$ in Cartesian geometry. This Cartesian approximation should fail only at small radii when the mass in cold gas exceeds the mass in hot gas,\footnote{Note that this is a more stringent condition than requiring the cloud/stream size to be small compared to other length scales, such as $r_{\rm stream}/r \ll 1$} $M_{\rm c}(<r) \gsim M_{\rm h}(<r)$, where the backreaction of the stream on the hot gas cannot be ignored. In this case the latter cannot be approximated as a quasi-infinite mass and momentum reservoir.
This is generally not an important consideration for the cloud cases, while crudely (and conservatively), this happens when $r \sim 0.1 R_{\rm vir}$ in the stream case. Using realistic potentials is an important extension of \citet{tan23-gravity}, which only simulated constant gravity for clouds, and used analytic methods to model the outcome of clouds falling in more realistic profiles. This also extends work in \cite{Aung2024} with realistic profiles for streams to include magnetic fields. We use two different setups for this configuration, at z~0 and at z~3 for clouds and streams, respectively. We describe these in detail when discussing the results.

Observational constraints on magnetic fields in the CGM are unfortunately currently rather sparse\footnote{For instance, the $B < 0.8 \, \mu{\rm G}$ constraint from Fast Radio Burst observations of \citet{Prochaska2019} gives a weak constraint, $\beta = P_{\rm gas}/P_{\rm B} \gsim 0.5 P_{\rm -3}$, where $P_{3} = P/(10^3 \, {\rm cm^{-3} \, K)}$ is the assumed CGM thermal pressure}, although we expect B-fields to have strengths between the $\beta \sim 1$ conditions of the ISM and $\beta \sim 100$ conditions of the ICM. We hence simulate weak to moderate fields corresponding to $\beta \sim 1 - 100$. We consider two magnetic geometries, with B-fields aligned with or orthogonal to the direction of gravity, respectively. These diametric cases are useful for illustrating the impact of B-field orientation. Again, while direct constraints on B-field geometry in the CGM are lacking, one might reasonably expect aligned fields to arise from strong inflows or outflows which `comb out' the B-fields radially, while (for instance) transverse fields might arise in the plane of Galactic disk. 

In our implementation of radiative cooling, we assume collisional ionization equilibrium (CIE) and solar metallicity ($X$ = 0.7,
$Z$ = 0.02). We obtain our cooling curve by performing a piece-wise power law fit to the cooling table given in \cite{GnatSternberg} over 40 logarithmically spaced temperature bins, starting from a temperature floor of $T \sim 10^4$ K, which we also enforce in the simulation. We then implement the fast and robust exact cooling algorithm described in \cite{Townsend}. For this cooling curve, the cooling time in the cold gas at $n=10^{-2} \, {\rm cm^{-3}}$ is $t_{\rm{cool}} \sim $ 0.15 Myr. As in \citet{tan23-gravity}, we also modify the cooling rate as
\begin{equation}
    \Lambda(T) = \Lambda_0\Lambda_{\rm{fid}}(T)
\end{equation}
where $\Lambda_0$ is the cooling strength so that $\Lambda_0=1$ corresponds to the fiducial cooling rate. In terms of cooling rates, modifying $\Lambda_0$ is equivalent to modifying the background pressure and density (or metallicity), and it allows us to explore the effects of different cooling times. To emulate the effect of heating and to prevent the background medium from cooling, we cut off any cooling above 5 $\times 10^5$ K. The exact choice of this value has been shown to be unimportant for both clouds and streams, provided it is high enough to not interfere with the cooling of mixed gas \citep{gronke18,mandelker20,Abruzzo2022}. In addition we define `cold' gas to be any material in the box that satisfies the threshold temperature condition $T <3\times10^4$ K.

\begin {table*}[t]
\caption {List of simulations and parameters.}
\label{table:sim_param}
\begin{center}
\begin{tabular}{c c c c c c c c} 
 \hline\hline
 Setup & Sim.      & Radius (pc) & $\Lambda_0$& $\beta$   & Field orientation & $g$ & Background\\
 \hline\hline
 Cloud+Stream & hyd        & (30,300,3000) & 1         & $\infty$     & - & const & Uniform+Strat.\\
 Cloud+Stream & hyd\_C30        & (30,300,3000) & 30        & $\infty$     & - & const & Uniform\\
 Cloud+Stream & hyd\_C100       & (30,300,3000) & 100      & $\infty$     & - & const & Uniform\\
 Cloud+Stream & mag        & (30,300,3000) & 1         & $1, 10, 100$     & aligned & const & Uniform+Strat.\\
 Cloud+Stream & mag\textunderscore C30    & (30,300,3000) &  30        & $1, 10, 100$ & aligned & const & Uniform\\
 Cloud+Stream & mag\textunderscore C100   & (30,300,3000) & 100       & $1, 10, 100$ & aligned & const & Uniform\\
 Cloud & mag\textunderscore hor  & (100, 300)  & 1     & $10, 100, 10^3, 3000, 10^4$ & transverse & const & Strat.\\
 Stream & mag\textunderscore hor  & (100, 300)  & 1     & $10, 100$ & transverse & const & Strat.\\
 Cloud+Stream & mag\textunderscore hor\textunderscore C30   &  300  & 30     & $10, 100$ & transverse & const & Strat.\\
 Cloud+Stream & mag\textunderscore hor\textunderscore C100    & 300 & 100     & $10, 100$ & transverse & const & Strat.\\
 \hline
 Cloud & hyd & 50 & 1 & $\infty$ & - & Hernquist & Hernquist\\
\hline
Stream & hyd & $10^3$, 3000, 5000, 20000 & 1 & $\infty$ & - & NFW & NFW \\
Stream & mag     & 3000 & 1 & $10, 100$  & aligned & NFW & NFW \\
Stream & mag\_hor    & 3000 & 1 & $10,100$  & transverse & NFW & NFW\\
\hline\hline
\end{tabular}
\end{center}
\end{table*}

The broad list and names of the simulations and the varying input parameters are presented in Table \ref{table:sim_param}.
For these background profiles, we run both cloud and stream simulations for radii $r = (30, 300, 3000)$ pc. We keep the number of cells across clouds and streams constant by scaling the box sizes by the same factor, and consider cooling strengths $\Lambda_0 = 1,30,100$.
The motivation for these choices is as follows. Infalling cloud survival in the hydrodynamic case is mostly sensitive to the cooling time (see \S 2.3 and Fig 2 of \citealt{tan23-gravity}). For our adopted gravity $g$, cloud overdensity $\delta \sim 100$ and pressure $nT\sim 100\ \rm{cm^{-3}K}$ (as adopted for for the uniform box), the critical cooling time for survival is $ t_{\rm cool} \sim 5 \times 10^{-3}$ Myr, to within a factor of 2 (due to the weak dependence of the survival ratio on the radius, i.e. $t_{\rm{grow}}/t_{\rm{cc}}\sim R^{-1/16}$) over the size range $3 \, {\rm pc}  < r < 3 \, {\rm kpc}$. Thus, our choice of $\Lambda_0 = 1,30,100$, corresponding to $t_{\rm cool} \sim 0.15, 5 \times 10^{-3}, 1.5 \times 10^{-3}\,$Myr, corresponds to cloud destruction, marginal survival, and definite survival and growth in the hydrodynamic case for. However, cloud size {\it does} affect the drag force and terminal velocity $v_{\rm T} \sim g t_{\rm grow}$ of clouds, since the growth time $t_{\rm grow} \sim m/\dot{m}$ is sensitive to cloud size. An important boundary is the transition from subsonic to supersonic infall, when the functional form of $t_{\rm grow}$ changes. When the shear becomes supersonic, the turbulent mixing velocity saturates and stops scaling with the cloud velocity. The cloud size where the terminal velocity transitions from subsonic to supersonic can be found using the criteria for subsonic infall $t_{\rm{grow}}<t_{\rm{ff}}$ where $t_{\rm{ff}}\sim c_{\rm{s,hot}}/g$, and substituting in the $t_{\rm{grow}}$ expression with $v=gt_{\rm{grow}}$. This yields a transition cloud radius for subsonic to supersonic terminal velocity as $r_{\rm sonic}\sim 150 \, \rm{pc}(t_{\rm cool}/0.03 \, {\rm Myr})^{-1/3}$, for our values of $g,\delta,c_{\rm s,hot}$. Thus, our cloud sizes of $30,300,3000$ pc correspond to subsonic, transonic, and highly supersonic infall. 

\section{Testing the hydrodynamic theory with a realistic setup}
\label{sec:test_theory}
\begin{figure*}
    \includegraphics[width=0.49\textwidth]{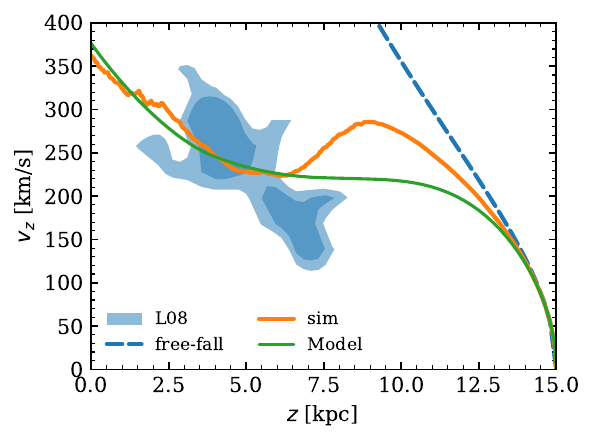}
    \includegraphics[width=0.48\textwidth]{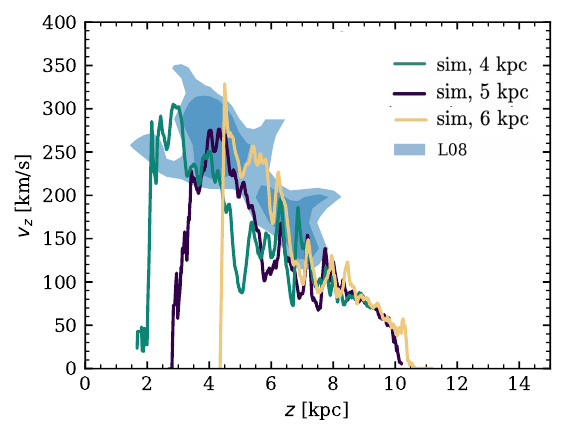}
    \caption{\textit{(Cloud, Realistic Background)}(Left) Velocity of the center of mass of cold gas for a cloud of radius 50 pc dropped from a height of $z=15$ kpc falling through an isothermal $T\sim 10^7$ K halo in hydrostatic equilibrium in a Hernquist gravitational potential. We also show the accretion model prediction (Section \ref{sec:theory}) and the free fall prediction along with the constraints of the observational velocity from \citealt{Lim2008} (L08) (shown as contours) for the Perseus cluster. (Right) Cold gas velocity profile as a function of z (averaged in x and y), at a specific instant in time when the center of mass has reached  $z=4,5,6$ kpc from the center of the galaxy. Note the excellent match with observations.}
    \label{fig:fc_real}
\end{figure*}
To validate the hydrodynamic theory presented in Section \ref{sec:theory}, before moving on to the MHD case, we simulate a cloud falling in a realistic setup. We consider cold clouds falling from $\sim 15\ \rm{kpc}$ toward the center of the cD galaxy NGC 1275 (Perseus A) which is in the Perseus cluster. The observed velocities of such infalling clouds are significantly lower than their ballistic free-fall velocities, and cannot be explained by the standard hydrodynamic drag model without fine-tuning \citep{Lim2008}. By applying the analytic models described in Section \ref{sec:theory}, \citet{tan23-gravity} showed that the slow infall velocities could be explained by accretion drag. We improve on their work by directly performing numerical simulations, and compare against results from the analytic model and the observations.

Following \cite{Lim2008}, we assume the Dark Matter halo has a Hernquist density and gravitational profiles \citep{Hernquist1990}:
\begin{equation}
    \rho(r) = \frac{M}{2\pi}\frac{a}{r(r+a)^3}
\end{equation}
\begin{equation}
    \phi(r) = -\frac{GM}{r+a}
\end{equation}
where $M=8.3\times10^{11}M_{\odot}$ is the total mass of the galaxy and $a=6.8$ kpc is the scale radius (numerical values from \citealt{Smith1990}). We assume an isothermal $T\sim 10^7$ K hot gas component in hydrostatic equilibrium with this gravitational potential (i.e. we ignore self-gravity), with a central gas density of $n=4\times10^{-2}\ \rm{cm^{-3}}$ \citep{Churazov2003}. We initialize the cloud with a radius of $50$ pc with an overdensity $\delta = 1000$ and track its evolution as it falls in a purely hydrodynamic simulation. Later on we shall see that the incorporation of magnetic fields do not significantly affect the cloud dynamics. 

Fig. \ref{fig:fc_real} shows the velocity evolution of the cold gas. The left plot shows the simulated velocity of the cold cloud center of mass, the model and free fall predictions, and the observed velocities of the cold filaments in the Perseus cluster \citep{Lim2008}. The simulation and model agree well\footnote{Note that there is an initial transient velocity overshoot in the simulation, which is not captured by the model. This is likely an effect of modelling the delay in the onset of turbulence by a constant factor $f_{\rm{kh}}$, when in reality this could be much more complicated. However, once turbulence sets in, the simulated velocities come back down to agree well with the model.}. Both agree well with the observed velocities, whereas the free-fall prediction is highly discrepant. \citet{tan23-gravity} showed that while free-fall predictions require strong fine-tuning of the drop height (i.e., the height at which the cloud starts falling) to match observations, the accretion braking model does not\footnote{A cloud that is dropped from very large distances can become destroyed before it manages to brake (due to weak cooling and drag forces at large radii), but the range of survival distances is considerably larger than for ballistic infall.}. While we only show one example here, our simulation results agree with these conclusions.

The right panel of Fig. \ref{fig:fc_real} shows cold gas velocity profiles when the cold gas center of mass is 4,5 and 6 kpc away from the galactic disk, along with the observational contours. These agree remarkably well, showing that our simulations do an excellent job at reproducing these observed velocity profiles. Importantly, the cloud does not reach a fixed terminal velocity for this hot gas background profile, in contrast with the constant gravity, stratified hydrostatic isothermal halo \citep{tan23-gravity}. 
This is because gravitational acceleration is not constant but continually increasing as the cloud falls inward. 

\section{Results: Magnetized Infalling Cold Clouds}
\label{sec:results_clouds}
\begin{figure*}
     \centering
         \includegraphics[width=0.49\textwidth]{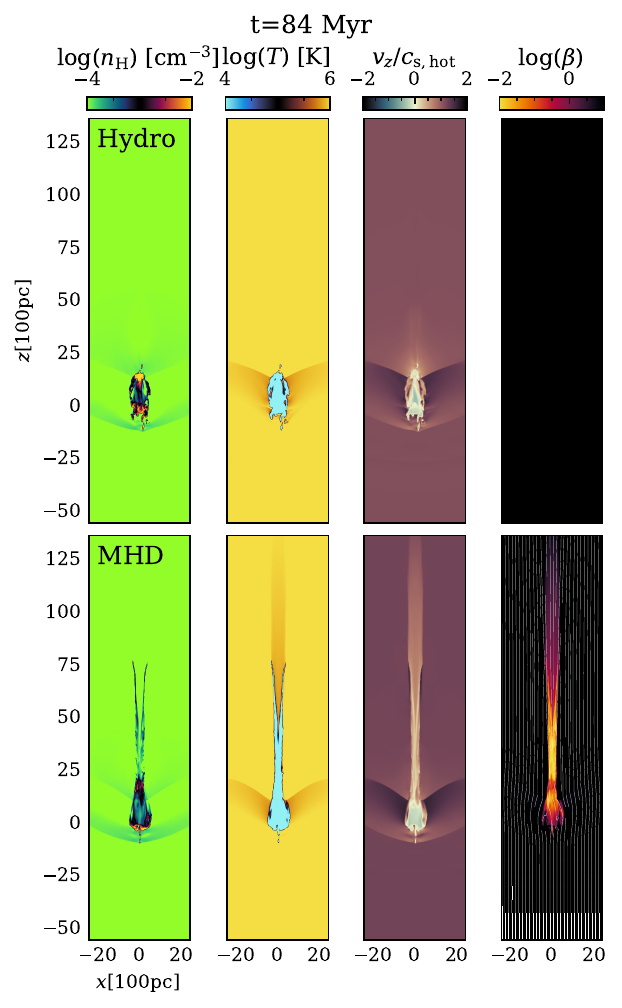}
         \includegraphics[width=0.49\textwidth]{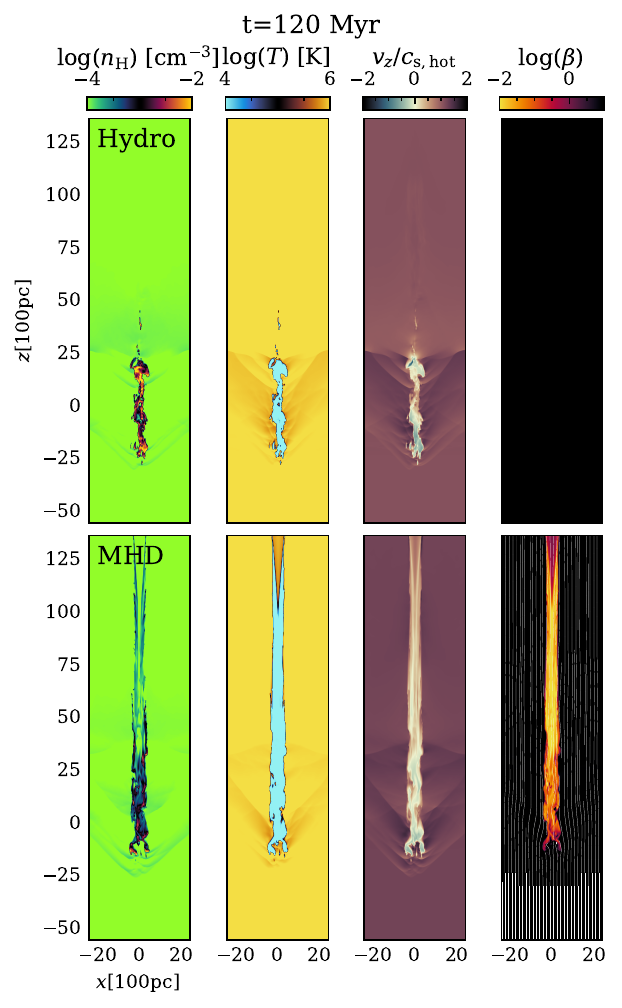}
        \caption{\textit{(Cloud, Uniform Background)} Snapshots of log-scaled quantities showing the time evolution of an infalling cloud with initial radius $r=300$ pc in a uniform background with and without a vertical (aligned) magnetic field. These simulations have $\Lambda_0=100$, and $\beta=10$ for the MHD case, corresponding to C100 and mag\textunderscore C100 cases respectively. The left and right panels are taken at $t=84$ Myr and $t=120$ Myr respectively while the top and bottom rows show simulations without and with aligned magnetic fields.  Each panel shows slices of log-scaled density, temperature, Mach number relative to the hot gas sound speed and plasma beta. The simulation times were chosen to reflect the initial disruption and the subsequent growth of the infalling cloud. 
        Since the hydrodynamic runs have no magnetic fields, the corresponding plasma beta slice is left blank.}
        \label{fig:fc_viz}
\end{figure*}
We now present the results of our simulations of magnetized infalling cold clouds. We start by considering a simplified environment with a uniform background. This allows us to investigate the impact of magnetic fields and compare it to the simulations and analytic models for the hydrodynamic case. We then perform the same analyses for simulations with a more realistic stratified background. In both cases, we vary the size of the clouds, the strength and orientation of the magnetic fields. We vary the cooling time in the uniform background case as well. Our goal is to study the impact of magnetic fields on cloud morphology, mass growth and survival.

Broadly speaking, when a cloud falls through a uniform background, we find that cloud survival and growth are similar between the MHD and hydrodynamic cases, despite evident morphological differences. For a cloud falling through a stratified background, the mass growth rate remains broadly similar with magnetic fields\footnote{More precisely, we find that aligned fields have little effect on mass growth rates, while transverse fields suppress mass growth rates, but this is compensated by a longer infall time, so total mass growth is enhanced.}, for those which already survive in the hydrodynamic case. However, magnetic fields can enhance survival over the hydrodynamic case. This effect becomes stronger for transverse fields and higher field strengths.
In fact, transverse B-fields enable cloud survival in every case we tested, due to magnetic tension in draped B-fields suppressing the KHI as the infalling cloud continually sweeps up field lines. This effect is considerably stronger than similar setups in wind tunnels (no external gravity), where B-fields can act to delay cloud destruction, but do not seem to change the size threshold for long-term cloud survival \citep{gronke20-cloud,hildalgo23}. The following sections discuss these findings in greater detail.

\subsection{Infalling Clouds: Uniform Background}
\subsubsection{Uniform Background: Overview}

\begin{figure}
    \centering
    \includegraphics[width=0.48\textwidth]{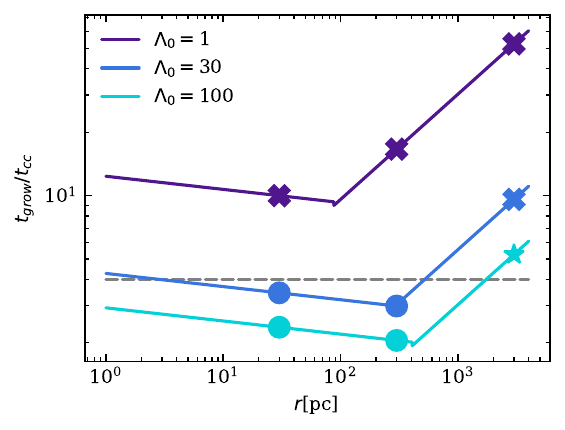}
    \caption{\textit{(Cloud, Uniform Background)} Cloud survival or destruction with magnetic field strength $\beta=10$ in a uniform background. We find identical results when $\beta=100$. The cooling strength parameter $\Lambda_0$ is varied. Crosses indicate a destroyed cloud, while circles indicate a survived and growing cloud. The star represents a case where it is unclear what the final state of the cloud is. The gray dashed line is the survival criteria from \citet{tan23-gravity} for hydrodynamic setups. The colored lines are analytic estimates of $t_{\rm grow}/t_{\rm cc}$ as a function of $r$ for different cooling times based on Section \ref{sec:theory}. The kink in these lines is where the terminal velocity transitions from subsonic to supersonic at large radii.}
    \label{fig:fc_survival}
\end{figure}

\begin{figure*}
     \centering
     
         \includegraphics[width=0.48\textwidth]{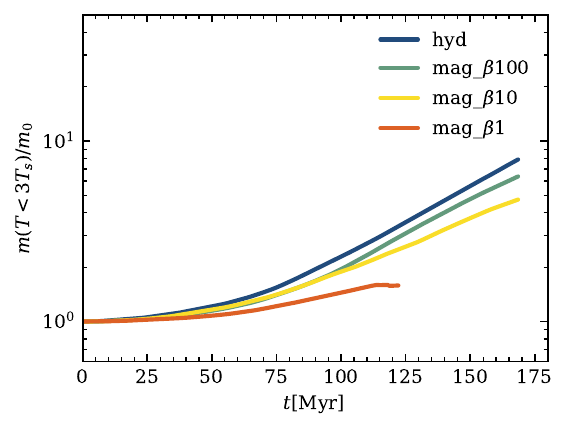}
         \includegraphics[width=0.48\textwidth]{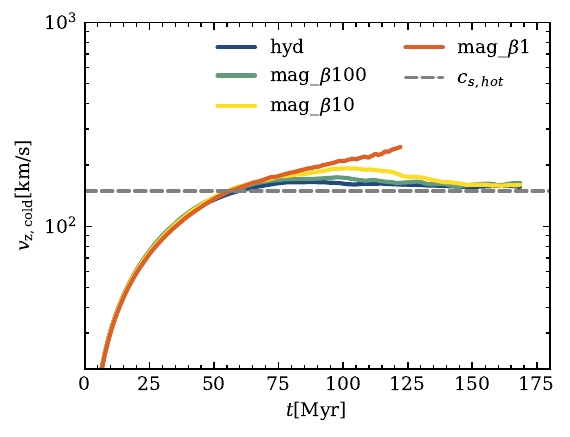}
         \includegraphics[width=0.48\textwidth]{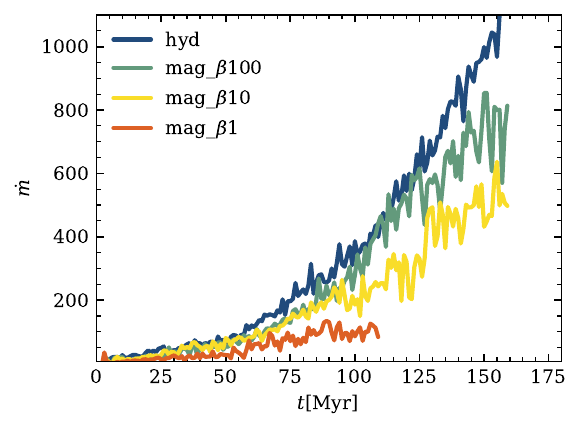}
         \includegraphics[width=0.48\textwidth]{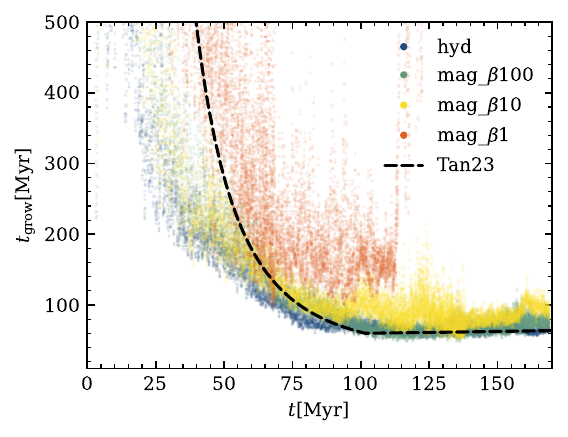}
              
    \caption{\textit{(Cloud, Uniform Background)} Results for the evolution of an infalling cold cloud of radius 300 pc in a uniform background with $\Lambda_0=100$ and varying plasma $\beta$. The suffix C100 is omitted here in the legend for brevity. (Left Top) The evolution of the cold gas mass as a function of time.  (Right Top) Velocity in the $z$ direction for the cold gas. (Bottom Left) The cold mass growth rate as a function of time. (Bottom Right) Comparison of the growth timescale $t_{\rm{grow}}$ along with hydrodynamic analytic theory shown as the black dashed line. The growth time increases weakly as $\beta$ is reduced.}
    \label{fig:fc_uniform_300pc}
\end{figure*}        

In this setup, we drop a cold and overdense cloud through an isothermal background of uniform density, both with and without magnetic fields.  We vary both the magnetic field strength and the cooling strength $\Lambda_0$. The hydrodynamic survival of the cloud is most sensitive to the cooling time (see end of Section~\ref{sec:numerical_methods}).

In this section, we only present results from simulations where the magnetic fields are aligned with the direction of gravity. When we discuss the more realistic case of stratified backgrounds, we will also show results from simulations with transversely aligned fields.
We found setups with transverse fields in a uniform background numerically challenging to run for sufficiently long times (beyond 50 Myr) to make meaningful statements. Such runs led to stalling simulations with prohibitively small timesteps. This is a common issue in magnetized multiphase simulations with short cooling times. For example \citet{gronnow22} noted a similar problem in their transverse field simulations, which they attributed to the fast magnetosonic speed $c_{ms}=\sqrt{v_A^2+c_{\mathrm{s,hot}}^2}$ dominating over $c_{\mathrm{s,hot}}$ in the CFL condition leading to extremely small timesteps required for stability. This can be traced to regions with strong magnetic fields and low density, which are clearly visible in the tails of magnetized clouds in the bottom row of Figure \ref{fig:fc_viz}.  We thus leave the discussion of the effect of transverse fields to the following section with a stratified isothermal hydrostatic background. There, because the background plasma $\beta$ in the transverse case increases as the cloud falls (i.e. gas pressure becomes increasingly dominant), these effects become less important.

We first discuss the qualitative differences with and without aligned magnetic fields. Fig. \ref{fig:fc_viz} shows slices of an infalling cold cloud of radius 300~pc in a uniform background in the $\Lambda_0=100$ case (where the hydrodynamic cloud is expected to survive and grow), for the hydrodynamic and $\beta=10$ cases with aligned B-fields. These snapshots show the density, temperature, velocity and magnetic field at two different times.
As the cloud falls, it develops a cometary tail, as seen in many previous wind-tunnel simulations. In this cooling regime, where $t_{\rm grow}/t_{\rm cc} = 2$ in the hydrodynamic case, the cloud is expected to survive and grow in mass since it is within previously established survival constraints ($t_{\rm{grow}}<4t_{\rm{cc}}$ see Section~\ref{sec:theory} for details). This is indeed what we observe for both the hydrodynamic and MHD cases. As the cloud falls, the hot phase mixes and cools in the forming tail. This enables the cloud to `survive' even if the initial head of cold gas ends up being disrupted. The infall is initially ballistic in nature, and the cloud accelerates to transonic velocities with respect to the hot gas. The resultant shocks are visible in the Mach number slice. From the plasma $\beta$ snapshots, we see evidence of strong B-field amplification (here, by a factor of $\sim 100$), which is consistent with results of magnetized turbulent mixing layer simulations \citep{ji19,TRMLwBfield} and magnetized wind tunnel cloud simulations \citep{gronke20-cloud,li20,Cottle2020}. This is also consistent with flux freezing amplification of the field when the hot gas cools and is compressed to higher densities. Since $\beta \ll 1$ in most of the cold gas volume, the cloud becomes magnetically supported, consistent with magnetized wind-tunnel simulations \citep{hildalgo23}. However, while most of the volume is magnetically supported, most of the mass is not, and has relatively weak magnetic fields. Towards the end of the simulation, the mass-weighted $\beta\sim 3-5$ while the volume-weighted $\beta \sim 0.2-1$. Thus, the cloud has a `core' of high $\beta$ gas where MHD forces are unimportant, and a `halo' of lower $\beta$ gas where MHD forces play an important role.

There are also morphological differences when magnetic fields are included. Importantly, the cloud tails are much longer with magnetic fields. This is because as the cloud develops this tail it becomes magnetic pressure dominated, as seen from the plasma $\beta$ snapshot. The gas thus traces a relatively strong aligned field which keeps the tail structure intact and prevents disruption. This effect is absent in the hydrodynamic case, where the continuous background shear causes the tail to mix away and get disrupted into the background hot gas. 
The B-field streamlines show some tangling and draping that develop when the cloud drags along the magnetic field as it falls. Magnetic tension somewhat suppresses Kelvin-Helmholtz and Rayleigh-Taylor instabilities. As a result, the cloud develops a smoother appearance than the purely hydrodynamic case, which is noticeably much more clumpy and fragmented. As we shall see in Section \ref{sec:strat_bkg}, this effect is much stronger with transverse fields.

\subsubsection{Uniform Background: Survival}
How do aligned magnetic fields affect cloud survival? At least for the range of plasma beta $\beta \sim 10 - 100$ we simulate, they seem to have very little effect. Fig. \ref{fig:fc_survival} shows the final status (survived or destroyed) of the MHD simulations for both $\beta=10$ and 100 (which give identical results) 
with different sizes ($r=30,300,3000$\, pc) and cooling parameters $\Lambda_0=1,30,100$, both of which affect the ratio $t_{\rm{grow}}/t_{\rm{cc}}$, the key dimensionless parameter that determines cloud survival in the hydrodynamic case. Our $\beta=1$ simulations did not run for a sufficiently long time to conclusively say anything about long term survival.
We define survival to be the case when the total cold gas mass in the box has either remained constant or increased by the end of the simulation (or when cold material starts leaving the box, whichever comes first). This criterion can be box size dependent but if the cold gas survives in our finite sized simulation boxes, it will survive in realistic scenarios (which typically have larger volumes). The colored lines in Fig. \ref{fig:fc_survival} are curves of constant $\Lambda_0$. The kinks correspond to the cloud radius above which the terminal velocity becomes supersonic (as opposed to subsonic before), such that turbulence saturates and $t_{\rm{grow}}$ takes on different scaling relations  (see Section~\ref{sec:theory} for further discussion).
Also plotted as a dashed line is the survival threshold $t_{\rm grow}/t_{\rm cc} < 4$ for the purely hydrodynamic case (Eq.~\ref{eqn:cloud_survival}) as found in \citet{tan23-gravity}. We find that this criteria still holds true for the MHD case\footnote{The case with the star has an undetermined final state, since for large clouds the cold gas starts to escape the box before its fate becomes clear.}, even though the cloud morphology is noticeably different. 

\subsubsection{Uniform Background: Cloud Properties}
Next, we examine how magnetic fields affect the evolution of cloud properties. Fig \ref{fig:fc_uniform_300pc} shows the evolution of an infalling cold cloud of radius $r=300$~pc. We compare simulations with $\Lambda_0 = 100$ so that we are firmly in the growth regime. The different simulations vary the magnetic field strength, with $\beta=1$, $\beta=10$, $\beta=100$ or $\beta=\inf$ (hydro; no magnetic fields). All of the parameters shown, i.e., mass, velocity along $z$, and the cold gas mass growth rate are volume weighted averages.  
Interestingly, the inclusion of magnetic fields seems to have a minimal effect on cloud growth rate or survival since the cold mass and emission evolution curves do not differ drastically from the hydrodynamic equivalents. 
For clouds that survive and grow, we find only very small differences in mass growth rates (corresponding to a decrease in $\dot{m}$) in the MHD case\footnote{Note that we stop the $\beta=1$ simulation at $t=110\ \rm{Myr}$ when cold gas starts leaving the box. In this extreme case, where the cloud infall starts with high magnetic pressure, the tails survive for even longer, causing this magnetic pressure supported gas to leave the top of the box sooner than the other cases.}. This is consistent with results from wind tunnel simulations such as \citet{gronke20-cloud} who find that magnetic fields also do not significantly alter cloud evolution. The black dashed line in the plot of $t_{\rm{grow}}$ corresponds to the analytic growth time for a purely hydrodynamic run with $\Lambda_0=100$ (Eq. \ref{eq:t_grow_cloud}). Since the black dashed line uses a simple expression to estimate the onset of turbulence ($w_{kh}$ in Section \ref{sec:theory}), the actual evolution of $t_{\rm{grow}}$ in the simulations differs somewhat initially but converges to the same values.
As shown in the plot, magnetic fields have minimal impact on $t_{\rm{grow}}$, for moderately weak fields, even though the cloud morphology seems quite different with magnetic fields. When the initial magnetic field is strong ($\beta=1$), the suppression of cooling seems to have important affects which include a reduced efficacy of the accretion braking mechnism (orange curve in velocity plot) and a factor of 2 higher $t_{\rm grow}$. Despite this, the infall is still far from free-fall. However, such strong magnetic fields are unlikely to be found in the CGM.

\subsubsection{Uniform Background: Turbulent Mixing}
\begin{figure*}
     \centering

         \includegraphics[width=0.45\textwidth]{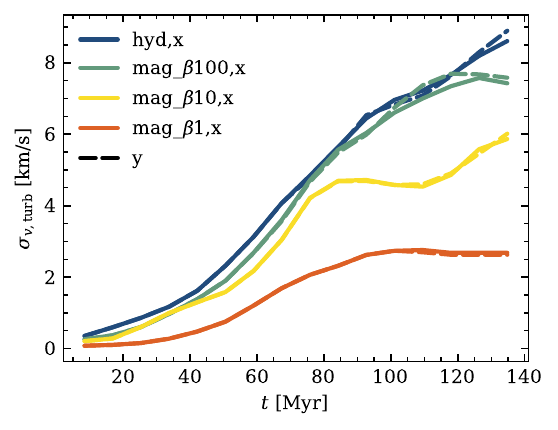}
         \includegraphics[width=0.45\textwidth]{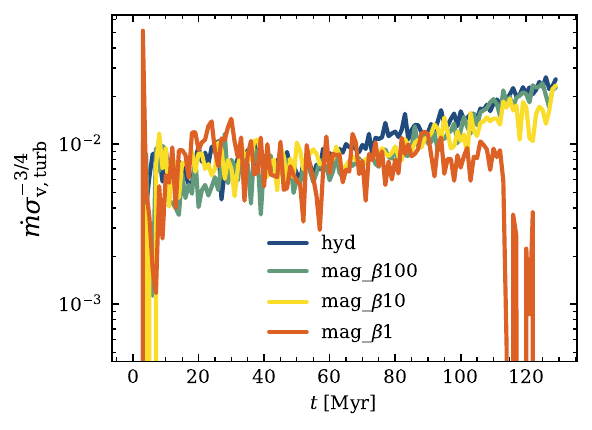}
        \caption{\textit{(Cloud, Uniform Background)} (Left) The turbulent velocity dispersion as a function of time in a growing cloud ($\Lambda_0=100$) for the hydrodynamic case and the aligned magnetic fields case with varying plasma $\beta$, and comparing the $x$ and $y$ components. (Right) The fairly constant value of $\dot{m}\sigma^{-3/4}_{\rm{v,turb}}$ shows that $\dot{m}\propto \sigma^{3/4}_{\rm{v,turb}}$, independent of $\beta$. The curve for the $\beta=1$ case should be ignored for $t>110$ Myr when cold gas starts leaving the box.}
        \label{fig:fc_mdot_comparison}
\end{figure*}      
The above results suggest that the inclusion of weak ($\beta \geq 10$) aligned magnetic fields has minimal impact on the evolution of infalling clouds. 
How do we connect this to the impact of magnetic fields on mixing?
In small scale simulations of turbulent radiative mixing layers with magnetic fields \citep{ji19, das23, TRMLwBfield}, instabilities are suppressed by magnetic tension, which leads to lower turbulent velocities, less mixing, and thus slower mass growth. These authors found that even when magnetic fields are too weak to suppress instabilities, they can still provide pressure support, lowering densities of the mixed phase and hence increasing cooling times, which can also lead to much slower mass growth rates. Similar effects take place for falling clouds, but they are much weaker. For instance, the $\beta=100$ falling cloud has almost identical mass growth rates as the hydrodynamic case, even though there is significant suppression of mass growth in the equivalent plane-parallel turbulent mixing layer simulations. For the latter, as B-field strength increases, the flow becomes nearly laminar.

The bottom left panel of Fig. \ref{fig:fc_uniform_300pc} shows the evolution of the mass growth rate for the purely hydro and aligned magnetic fields cases (with varying $\beta$). While small in magnitude, there is still a general decrease in $\dot{m}$ for stronger magnetic fields. 
What is the source of this trend?
Since one reason for a lower $\dot{m}$ in magnetized TRML simulations is the suppression of instabilities and hence $v_{\rm{turb}}$, we can directly check if the turbulent velocities are similarly smaller in the magnetized infalling clouds. 

We find that the difference in mass growth rates with and without magnetic fields arises directly from the different levels of turbulence. In the turbulent mixing layer setup, magnetic stresses enhance momentum transfer and reduce shear between hot and cold phases. Shear-induced turbulence is therefore greatly reduced. By contrast, in falling clouds, shear (and hence turbulence) is constantly driven by infall due to gravitational forces. The left panel of Fig. \ref{fig:fc_mdot_comparison} shows that the turbulent velocity dispersion (calculated as the standard deviation of turbulent velocities along either the x or y axis; these are found to be equal) is suppressed by a factor of $\sim 2-3$ as magnetic field strength increases. Moreover, in the right panel of Fig. \ref{fig:fc_mdot_comparison} we see that $\dot{m}\propto\sigma_{\rm{v,turb}}^{3/4}$ in both hydrodynamic and MHD cases, independent of plasma $\beta$. Thus, the reduction in $\dot{m}$ as B-field strengths increase is almost entirely explained by reduced mixing due to lower turbulent velocities; any change in cooling due to magnetic pressure support appears to be secondary.

\begin{figure*}
    \centering
    \includegraphics[width=0.49\linewidth]{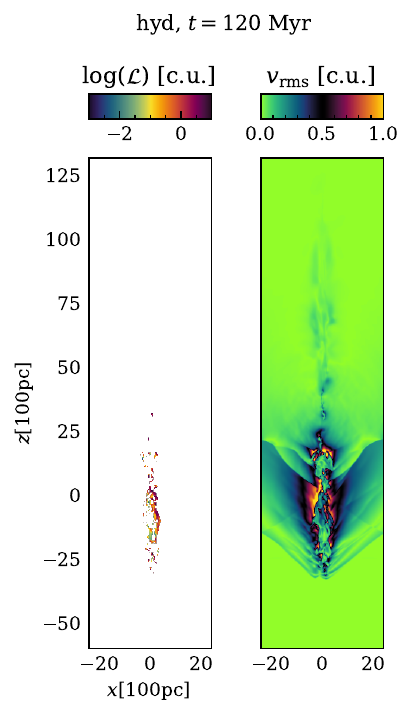}
    \includegraphics[width=0.49\linewidth]{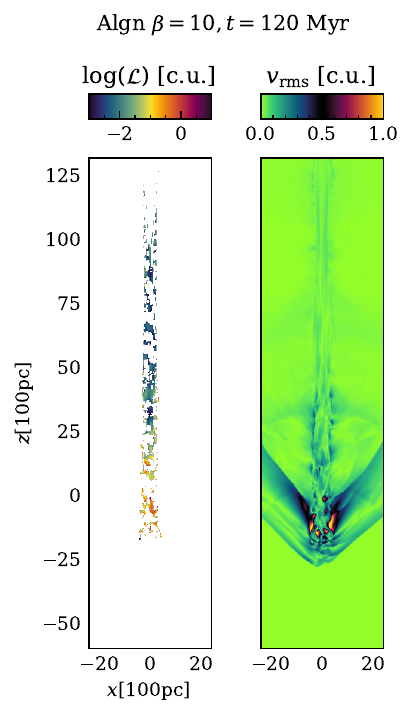}
    \caption{\textit{(Cloud, Uniform Background)} Maps of log-scaled luminosity and rms turbulent velocity for the late time ($t=120$ Myr) snapshot of the hydro and aligned magnetic fields with $\beta=10$ simulations, all in code units (c.u). White space in luminosity corresponds to regions where there is no emission. Overall, magnetic fields suppress turbulent mixing and cooling.}
    \label{fig:fc_ptherm1}
\end{figure*}

In addition, there are significant differences in \textit{where} cooling and condensation take place, between the hydrodynamic and MHD cases, as shown by Fig. \ref{fig:fc_ptherm1}. In the hydrodynamic case, cooling and condensation occur throughout the tail. However, in the MHD case, cooling is strongest at the head of the cloud; most of the cooling luminosity in the tail is almost two orders of magnitude smaller (although as noted above, most of the mass of cold gas is not volume filling). Note that this difference is also because the MHD case has more extended tails than the hydro case, and the regions where these is much less cooling in the MHD tails are regions where there is no tail in the hydro case. Despite this difference in cooling morphology, the hydrodynamic and MHD cases have almost identical mass growth rates. We will further investigate this interesting puzzle in future work.

\begin{figure}
    \centering
    \includegraphics[width=\linewidth]{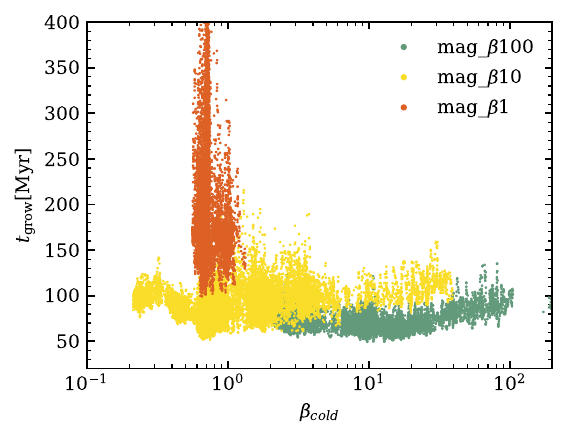}
    \caption{\textit{(Cloud, Uniform Background)} The independence of the evolution of the cold mass growth time $t_{\rm grow}$ and the plasma $\beta$ parameter in the cold gas for aligned magnetic fields with varying initial $\beta$ values. These points are taken when the growth time $t_{\rm{grow}}$ reaches a steady state for each respective case, to avoid the initial transient/growth phase. The large spread in $t_{\rm grow}$ for the $\beta \sim 1$ case is because the cloud evolves at fixed $\beta_{\rm cloud} \sim 1$, and the cloud has widely varying $t_{\rm grow}$ at different stages in its evolution.}
    \label{fig:tgrow_beta}
\end{figure}

\subsubsection{Uniform Background: Summary}

In summary we find that in a uniform background, cloud infall remains largely unaffected by the inclusion of aligned magnetic fields, for $\beta > 1$. The mass growth timescale $t_{\rm grow}$ is largely independent of $\beta$ unless the magnetic pressure come close to equipartition $\beta \sim 1$ (see Fig \ref{fig:tgrow_beta}), when $\dot{m}$ is slightly reduced, due to reduced mixing. 

\subsection{Stratified Background}
\label{sec:strat_bkg}
\subsubsection{Stratified Background: Overview}

Moving on from the simplified uniform background setup, we turn to the more realistic case of a stratified background. Here, the cooling time shortens as the cloud falls since $t_{\rm{cool}}\propto n^{-1}$ and the background density increases exponentially with height fallen. For fields aligned with the direction of gravity, $B=$ const and hence $\beta$ increases as the cloud falls. For fields transverse to the direction of gravity, we initialize the magnetic pressure to have the same exponential profile so that $\beta=$ const. 
We do not boost the cooling for the stratified case and always maintain $\Lambda_0 = 1$. 

\subsubsection{Stratified Background: Cloud Properties}
\begin{figure*}
         \centering
         \includegraphics[width=0.49\textwidth]{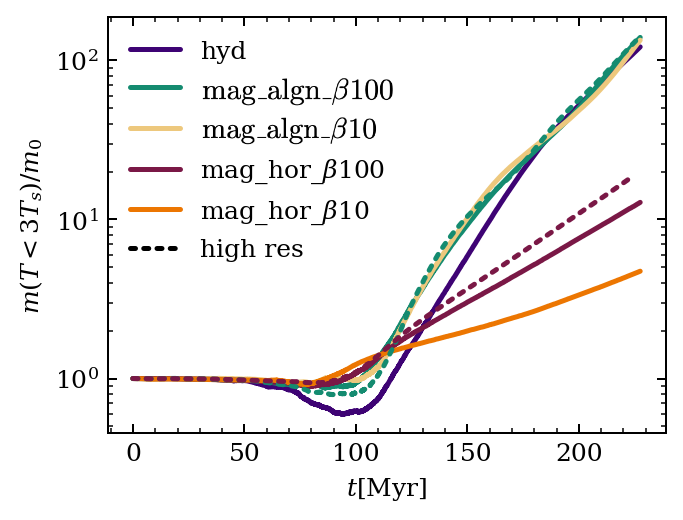}
         \includegraphics[width=0.48\textwidth]{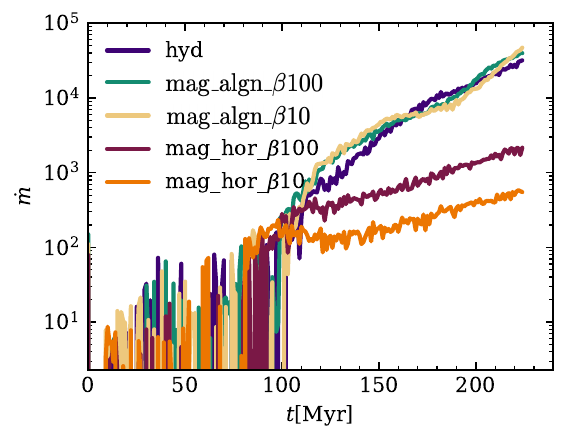}
         \includegraphics[width=0.48\textwidth]{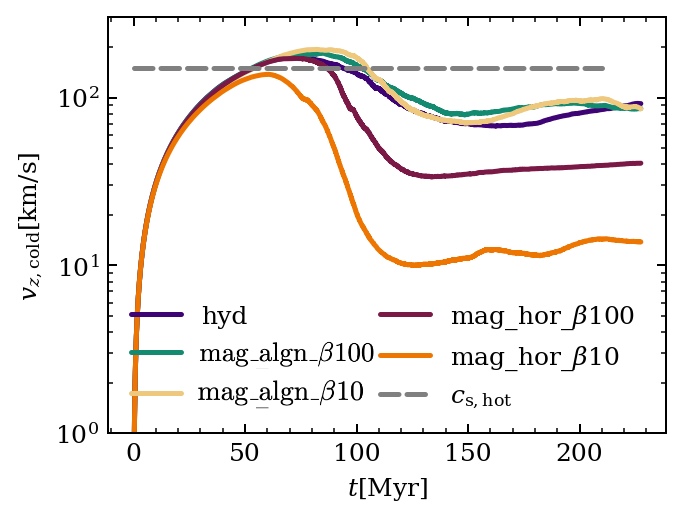}
         \includegraphics[width=0.48\textwidth]{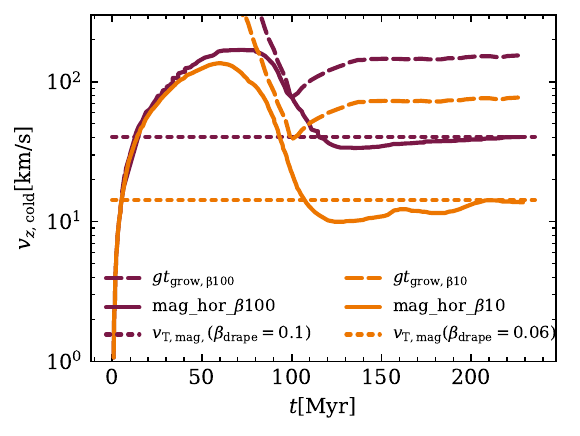}
         \includegraphics[width=0.48\textwidth]{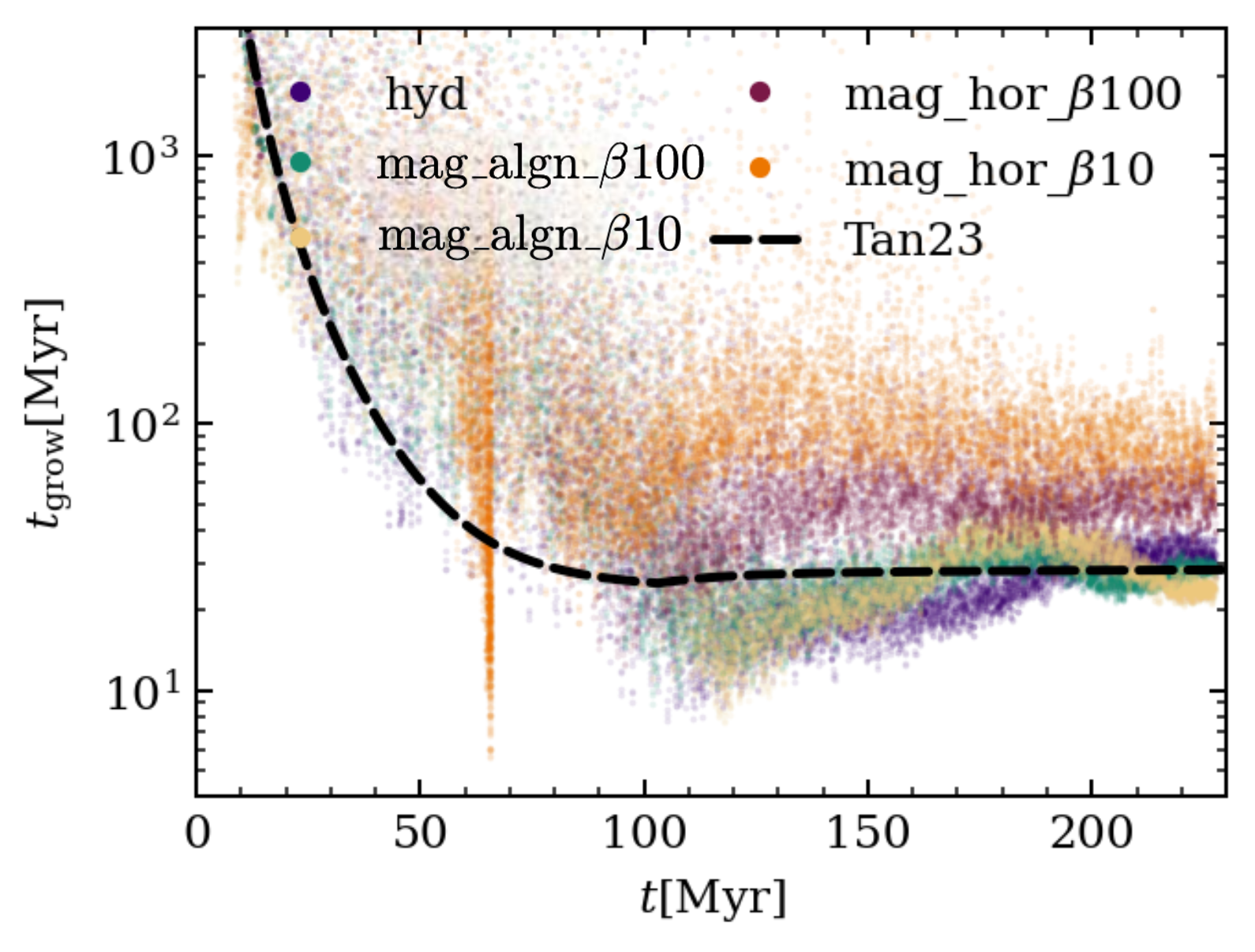}
         \includegraphics[width=0.48\textwidth]{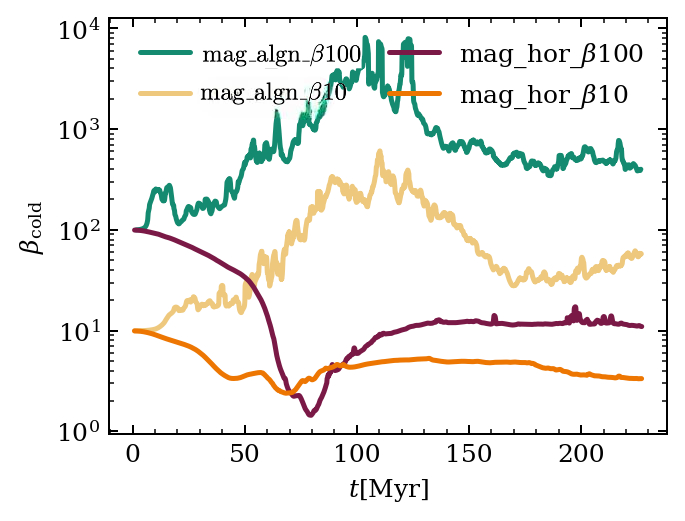}
        \caption{\textit{(Cloud, Stratified Background)} Results for the evolution of an infalling cold cloud of radius 300pc in an isothermal, constant gravity stratified background with the fiducial cooling strength, $\Lambda_0 = 1$. (Top left) The evolution of the cold gas mass as a function of time.  (Top Right) Mass growth rates. (Middle left) Velocity in the z direction for the cold gas. (Middle right) Comparison of the terminal velocity due to cooling accretion braking ($v_T=gt_{\rm grow}$) and due to magnetic braking ($v_{\rm T,mag}$ from Eq. \ref{eqn:v_T_mag}) for the example cases of cloud infall through transverse magnetic fields with $\beta_0=10, 100$. Here $\beta_{\rm drape}$ are taken to be typical values in the draping layer from the simulations. (Bottom left) Comparison of the growth timescale $t_{\rm{grow}}$ for the cases where the cloud survives along with analytic theory for the hydrodynamic case. (Bottom right) Evolution of the plasma $\beta$ in the cold gas. Note the non-monotonic evolution of plasma $\beta$, for reasons described in the text.}
        \label{fig:fc_strat_300pc_vz_t_grow}
\end{figure*}

Fig. \ref{fig:fc_strat_300pc_vz_t_grow} shows the evolution for infalling clouds in a stratified background. 
Since the background density increases as the clouds fall, the cold gas mass increases substantially (upper left panel), much more than the uniform background case from the previous section.  
When the cloud has aligned magnetic fields, there is almost no deviation from the non-magnetized case in $m(t)$ or $\dot{m}(t)$ (upper panels). This is consistent with our finding from the uniform background case that aligned fields have a minimal impact on the mass growth. As seen from the lower left panel for $t_{\rm grow}$, the rate of mass growth with aligned fields is roughly consistent with the analytic model for the hydrodynamic case. Since turbulence takes a while to develop and induce mixing and accretion drag, the initially free-falling cloud first overshoots in velocity as it falls into the dense regions, where accretion drag is strongest, before settling down to an asymptotic terminal velocity with fixed $t_{\rm grow}$. 

It may seem surprising that the cloud reaches a terminal velocity with constant $t_{\rm grow}$, given that it is continually falling into denser gas with a shorter cooling time, and stronger accretion drag force. As explained in \citet{tan23-gravity}, this occurs because of a feedback loop: if $t_{\rm grow}$ is large, the cloud will fall faster into denser regions, reducing $t_{\rm grow}$. As $t_{\rm grow}$ decreases, since it falls through into a denser and denser background, $t_{\rm grow}$ naturally increases due to the increase in cold gas mass from the mass dependence (see equation \ref{eq:t_grow_cloud}). One can show that $\dot{t}_{\rm grow}, \dot{v} \rightarrow 0$ on similar timescales (see equations 40,41 in \citealt{tan23-gravity}), so that $t_{\rm grow}, v \rightarrow$ const.  

\begin{figure*}
     \centering
     
        \includegraphics[width=0.48\linewidth]{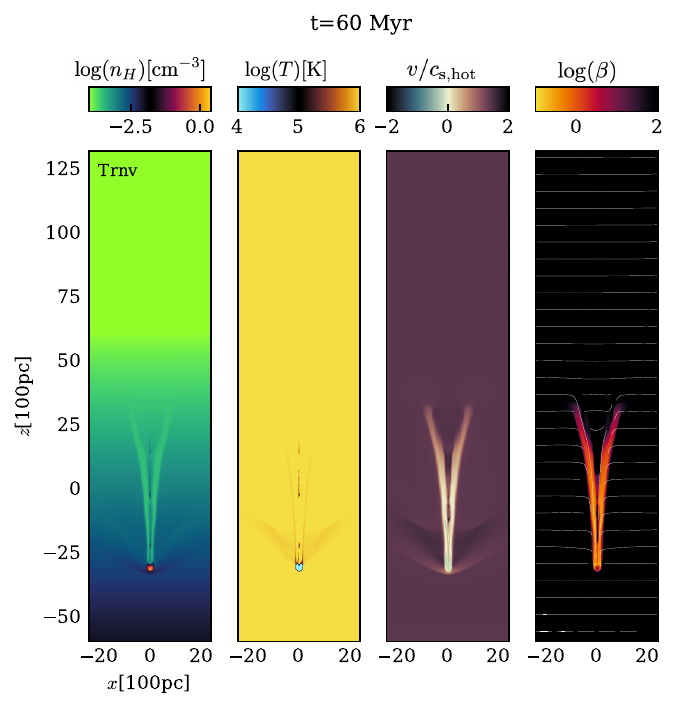}
        \includegraphics[width=0.48\linewidth]{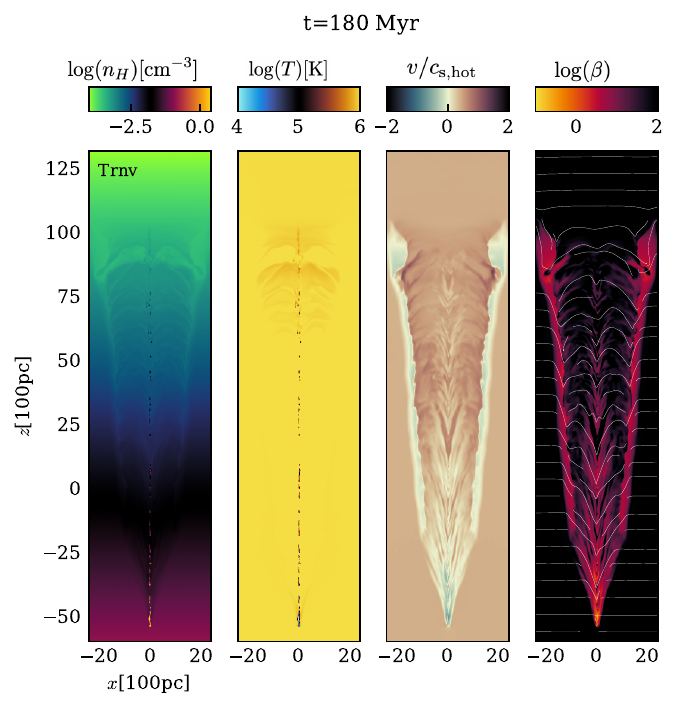}
    
        \caption{\textit{(Cloud, Stratified Background)} Snapshots showing log-scaled quantities of the time evolution of an infalling cloud with initial radius $r=300$ pc in a stratified background with transverse magnetic fields. The left and right panels are taken at $t=60$ Myr and $t=180$ Myr respectively.  Each panel shows slices of log-scaled density, temperature, Mach number relative to the hot gas sound speed and plasma beta.}
        \label{fig:fc_strat_viz_trnv}
        
\end{figure*}

With this setup, we are also now able to investigate the impact of horizontal fields. 
We find that mass growth is significantly suppressed for horizontal fields, compared to the hydrodynamic and vertical field case. The suppression is larger for stronger B-fields. Despite the lower $\dot{m}$, the terminal velocity $v_{\rm T}$ is {\it lower} in the horizontal field case, and continually decreases for stronger fields. This means that accretion drag is {\it not} the dominant drag force. Otherwise, the terminal velocity would be higher than the hydrodynamic case, given the longer mass growth times $t_{\rm grow}$. Indeed, in this case, magnetic drag appears to be dominant. Magnetic draping amplifies the B-field around the cloud to rough equipartition with ram pressure, as seen from the low $\beta$ regions around the cloud in Fig. \ref{fig:fc_strat_viz_trnv}. This strong magnetic field inhibits mixing via magnetic tension, and lowers gas overdensities and cooling rates via magnetic pressure support. This leads to an overall drop in $\dot{m}$; as we shall see, it also has implications for cloud survival. Lateral confinement by the strong magnetic field causes the cloud to become compressed in the direction orthogonal to infall, until it develops a thin vertical core of high density. At the same time, magnetic drag due to the strong field acts to brake the cloud's infall. For the cases we have simulated, this becomes more important than accretion drag. Magnetic drag enhances standard hydrodynamic drag by a factor of $[1+ (v_{\rm A}/v)^2]$, as in equation \ref{eqn:Fdrag}, so it can out-compete accretion drag $\dot{m} v$.  
Once magnetic drag dominates, we obtain a new terminal velocity $v_{\rm T, mag}$ that balances gravity as in Eq. \ref{eqn:v_T_mag}. We directly compare the two terminal velocities, $v_{\rm T,grow}=gt_{\rm grow}$ and $v_{\rm T, mag}$, in the middle right panel of Fig. \ref{fig:fc_strat_300pc_vz_t_grow}. This shows that for the infalling cloud with transverse fields with $\beta_0=10$ and 100 the terminal velocity due to magnetic drag is much lower and thus \textit{is} the cloud terminal velocity due to the low $\beta$ (and thus low Alfven Mach number $\mathcal{M}_A$) reached in the draping layer. 


The bottom right panel of Fig. \ref{fig:fc_strat_300pc_vz_t_grow} shows the evolution of plasma $\beta$ for the cold cloud gas. 
They show non-monotonic evolution for both vertical and horizontal fields. For vertical fields, the plasma $\beta$ first rises as the cloud mixes with higher $\beta$ gas (recall that $\beta$ of the background hot gas increases in the vertical field case). Then, as cooling becomes efficient and the cold gas mass starts to grow at $\sim 100$Myr, compressional amplification of B-fields in cooling gas decreases $\beta$. For horizontal fields, $\beta$ of the background gas is constant. Initially, $\beta$ falls due to compressional amplification of the draped field; the fields are strongly amplified during `velocity overshoot', when ram pressure is particularly high. Once the cloud settles to a lower terminal velocity, with a lower (constant) ram pressure, the draped field is weaker and $\beta$ settles at a higher, constant value.

\begin{figure*}
    \centering
    \includegraphics[width=0.49\textwidth]{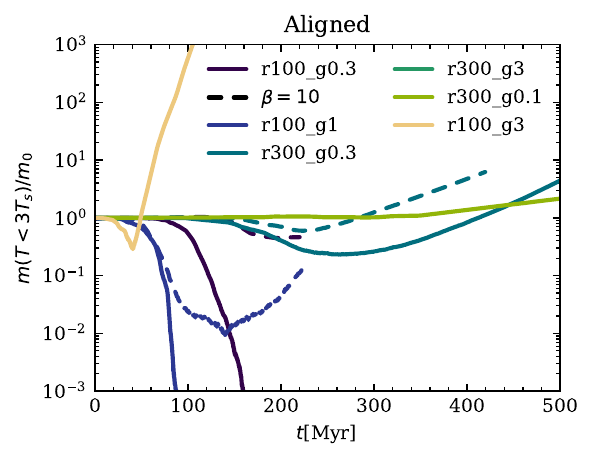}
    \includegraphics[width=0.49\textwidth]{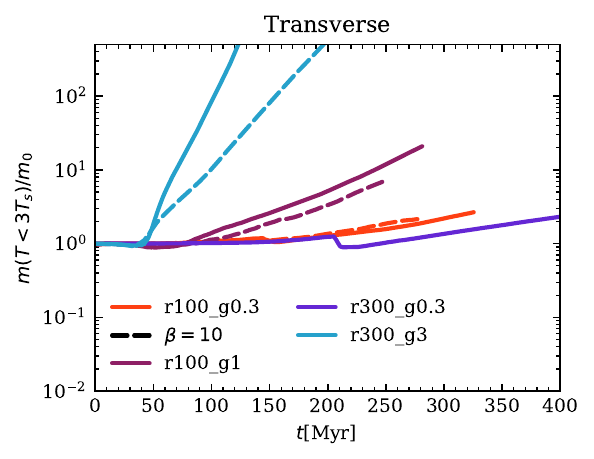}
    \caption{\textit{(Cloud, Stratified Background)} Cold mass evolution of clouds in a stratified background with (left) aligned magnetic fields and (right) transverse magnetic fields, for a range of cloud sizes and gravitational acceleration strengths in units of $g_{\rm{fid}}$ for the stratified medium. Note the different radii (in pc) prefixed with `r' and different gravitational strengths (in units of the fiducial acceleration, $10^{-8} \rm{cm^2/s}$) prefixed with `g' in the legends. Solid lines denote an initial $\beta=100$, while dashed lines denote an initial $\beta=10$. As before, the background $\beta$ increases with distance along the direction of cloud motion (and thus with time) for aligned fields, but stays constant for transverse fields.}
    \label{fig:fc_strat_full}
\end{figure*}

\subsubsection{Stratified Background: Resolution}
We check that our results for the stratified clouds with either aligned or horizontal magnetic fields are not significantly affected by resolution just as in the hydrodynamic case. The top left panel of Fig.~\ref{fig:fc_strat_300pc_vz_t_grow} includes mass growth histories of the simulations with $\beta=100$ at twice the fiducial resolution. We find that the higher resolution runs have a very similar mass growth history to the fiducial resolution.

\subsubsection{Stratified Background: Survival}
We can also ask if the survival criteria remains the same in the stratified case. To reiterate, we define survival to be the case where cold gas is left within the box and the total cold gas mass is increasing by the end of the simulation (if cold gas starts to escape the box, we stop the simulation early). 
Fig. \ref{fig:fc_strat_full} shows the evolution of the cold cloud with aligned and transverse magnetic field cases for varying sizes and gravitational strengths. The dashed and solid lines correspond to $\beta=10,100$ respectively. 

\begin{figure*}
    \centering
    \includegraphics[width=0.49\textwidth]{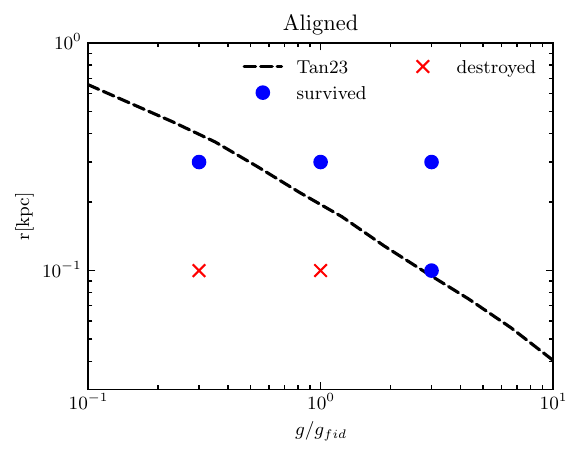}
    \includegraphics[width=0.49\textwidth]{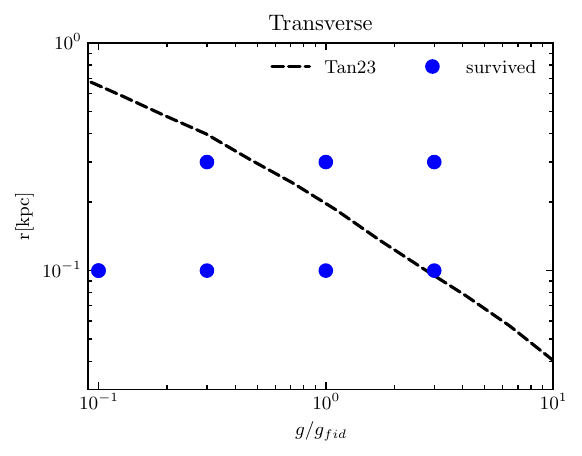}
    
    \caption{\textit{(Cloud, Stratified Background)} Final states of the cold  clouds in (left) aligned magnetic fields and (right) transverse magnetic fields with $\beta=100$ along with the survival criteria given by \cite{tan23-gravity} for the purely hydrodynamic infalling clouds in the same stratified background medium (dashed line). The 100 pc simulation runs, while not mentioned in the table of simulations, are added here to directly compare to the \cite{tan23-gravity} figure.}
    \label{fig:fc_strat_surv}
\end{figure*}

Survival criteria are summarized in Fig. \ref{fig:fc_strat_surv}, which should be compared against the mass evolution plots in Fig. \ref{fig:fc_strat_full}. For aligned fields, the $\beta=100$ survival criteria are similar to the hydrodynamic case (equation \ref{eqn:cloud_survival}, $t_{\rm grow} < f_{\rm S} t_{\rm cc}$). However, from Fig. \ref{fig:fc_strat_full}, we see that clouds which are destroyed when $\beta=100$ in fact survive and eventually grow when $\beta=10$. This enhancement of survival in the presence of magnetic fields is consistent with wind tunnel simulations (\cite{Pineda2023}), and it is a stronger effect than for the uniform backgrounds simulation. By contrast, for transverse fields, {\it all} clouds survive, over the range of parameter space that we scanned; the standard hydrodynamic survival criterion does not apply. Magnetic draping evidently strongly promotes cloud survival. This does not mean that transverse fields will always protect clouds, as we have only scanned a limited portion of parameter space. In wind tunnel simulations, clouds can still be destroyed in the presence of transverse fields \citep{Pineda2023}. We suspect that in this stratified case, draped fields delay mass loss until clouds 
fall to denser regions where the cooling time is sufficiently short that $t_{\rm grow} < f_{\rm S} t_{\rm cc}$. Unfortunately, given that we were unable to implement transverse fields in the uniform background case, where this effect would be absent, we cannot test this assertion. 
\subsubsection{Stratified Background: Transverse magnetic fields enhance growth and survival}
\label{subsec:clouds_transverse}

\begin{figure*}
    \centering
    \includegraphics[width=\linewidth]{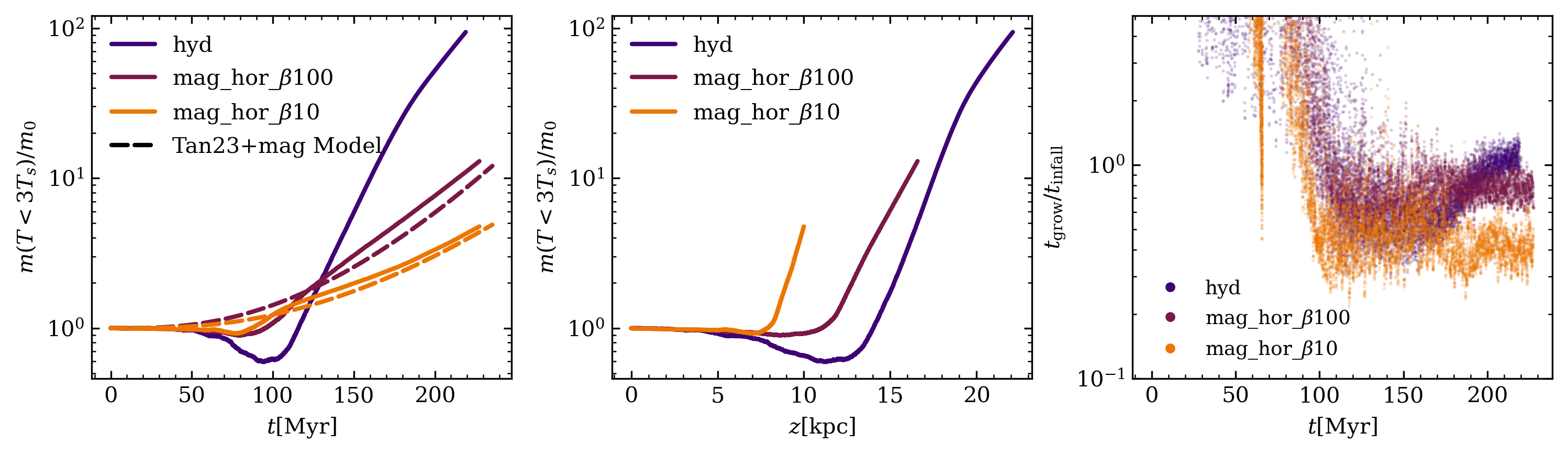}
    \caption{\textit{(Cloud, Stratified Background)}(Left) The evolution of cold gas mass with time along with the analytic model prediction that includes the magnetic drag term in addition to the hydrodynamic model from \citealt{tan23-gravity} (Eq. \ref{eqn:momentum_with_mag}). (Middle) The evolution of cold gas mass with height for the hydro and transverse fields cases showing enhanced growth with stronger transverse fields as compared to the hydrodynamical case. (Right) The ratio of the growth time to the infall time (defined as the time for the cloud to fall through a scale height $H$) for the same set of simulations showing the enhanced growth with transverse fields occurring due to the cloud spending more time falling through the same medium which increases the amount of cold gas accreted. }
    \label{fig:fc_strat_mag_hor_vs_z}
\end{figure*}

\begin{figure}
    \centering
    \includegraphics[width=\linewidth]{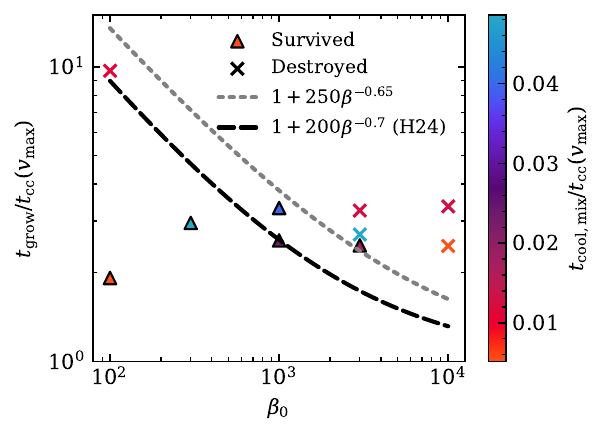}
    \caption{\textit{(Cloud, Stratified Background)} 
    The survival criteria for transverse magnetic fields with increasing initial plasma $\beta$. The y axis shows the ratio of growth time to the cloud crushing time evaluated at the peak velocity. Triangles denote survived clouds and crosses denote destroyed ones at the end of the simulation. The triangles and crosses are colored by the wind tunnel survival criteria ratio, $t_{\rm cool, mix}/t_{\rm cc}$ evaluated at $v_{\rm max},$ showing that this criteria fails to predict survival here. The gray dashed line denotes the inferred survival criteria, which is similar to the black dashed line from the \cite{hildalgo23} best fit criteria for clouds in a wind tunnel with transverse fields (replacing $t_{\rm cool,mix}$ for wind tunnels with $t_{\rm grow}$ for the infalling case).}
    \label{fig:fc_strat_mag_hor_survival}
\end{figure}

The left panel of Fig. \ref{fig:fc_strat_mag_hor_vs_z} shows that transverse magnetic fields suppress mass growth as one might expect, since magnetic tension inhibits mixing. However, the middle panel of Fig. \ref{fig:fc_strat_mag_hor_vs_z} shows that overall mass growth is actually \textit{enhanced} when cloud mass is plotted as a function of height. A cloud gains more mass overall when dropped from a given height, even if the rate of mass growth is slower. Additionally, reducing the plasma $\beta$ increases this effect. The reason for this is revealed in the right panel in the same figure. Here we plot the ratio of the growth time to the infall time (defined as the time for the cloud to fall through a scale height H) as a function of time. This ratio roughly decreases with decreasing $\beta$ indicating that the cloud slows down much more due to magnetic drag as $\beta$ decreases, and thus spends more time accreting cold gas from cooling for the same height. This behavior is well captured by adding in the magnetic drag term to the hydrodynamic model as shown from the left panel of the same figure. This naturally has consequences for survival. As shown in Fig. \ref{fig:fc_strat_surv}, the survival criteria here becomes non-trivial and does not follow the hydrodynamic infall criteria found in \citet{tan23-gravity}. Instead, the reason for enhanced survival is the draping of magnetic fields that provide an additional drag force to slow down the cloud, leading to more time for cold gas growth. The cloud however first goes through a velocity `overshoot' phase as previously described, where the fields get compressionally amplified to their maximum value at the peak of this overshoot. After this the cloud velocity reduces but the surrounding draped magnetic field retains its lowered $\beta_{\rm drape}$ value. Note that $\beta_{\rm drape}\ll\beta_{\rm cold}$, because the draping layer does not encompass the entire cold cloud. Consequently, the Alfven mach number $\mathcal{M_A}=v/v_{A}$ becomes low enough for the magnetic drag term to become important. Ultimately, this is responsible for the enhanced survival of clouds that should have seemingly gotten destroyed by the hydrodynamic criteria. Fig. \ref{fig:fc_strat_mag_hor_survival} shows the survival and destruction of clouds in terms of the ratio $t_{\rm grow}/t_{\rm cc}$ and the initial plasma $\beta$ parameter. The survival criteria from \cite{hildalgo23} for wind tunnel simulations is also plotted, where the term $t_{\rm cool, mix}$ is replaced by $t_{\rm grow}$ here. In addition,  we plot the slightly modified inferred criterion $t_{\rm grow}/t_{\rm cc} < 1+250\beta^{-0.65}$ that somewhat better fits the results.

\subsubsection{Stratified Background: Summary}

In summary, vertical magnetic fields aligned with gravity do not seem to have significant effects. Mass growth rates and survival criteria are similar to the hydrodynamic case, although we do see enhanced survival for stronger fields. By contrast, transverse fields have significant effects due to magnetic draping. Magnetic tension reduces mixing but interestingly increases mass growth as a function of height, since strong magnetic drag significantly reduces terminal velocities. At the same time, transverse fields strongly enhance survival. Thus, the fate of infalling gas clouds can depend strongly on B-field orientation. 

\section{Results: Magnetized Infalling Cold Streams }
\label{sec:results_streams}

We now describe results for infalling cold streams, initially with a constant gravitational field (in both uniform and stratified backgrounds), and later in a more realistic potential. As we shall see, the inclusion of gravitational forces introduces important differences from the standard wind tunnel case. It is important to keep in mind that while wind tunnel simulations are justified for clouds which are being entrained in a galactic wind, cold streams {\it always} fall under the influence of gravity. We describe the evolution from the standpoint of morphology, stream growth and survival, as well as the role of magnetic fields in stream evolution. The different geometry of cold streams results in important differences from infalling clouds. 


\subsection{Uniform Background: Overview}
\begin{figure*}
     \centering
     
         \includegraphics[width=0.48\textwidth]{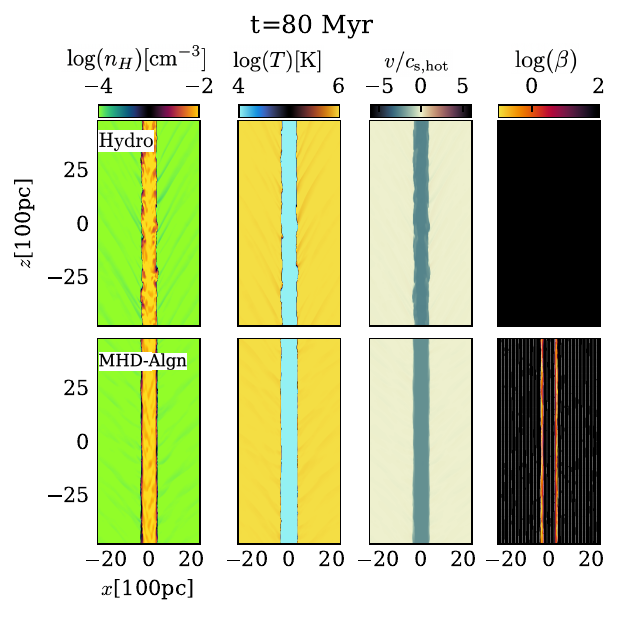}
         \label{fig:cs_viz1}
         \includegraphics[width=0.48\textwidth]{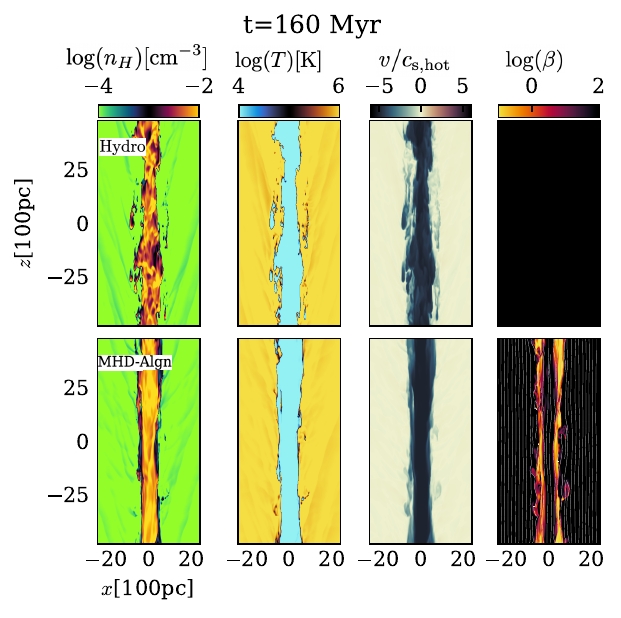}
         \label{fig:cs_viz2}
     
        \caption{\textit{(Stream, Uniform Background)} Evolution snapshots of the stream of radius $r=300$ pc falling under gravity in a uniform background showing slices of density (first column), gas temperature(second column), velocity in the z direction in units of the hot gas sound speed (third column) and plasma $\beta$ (fourth column) for the purely hydro case (top row) and aligned B fields case with $\beta=100$(bottom row) at $t=80$ Myr (left panel) and $t=160$ Myr (right panel). These times were chosen to reflect the initial disruption from KHI and the subsequent growth due to radiative cooling in the middle of the evolution of the falling stream. These runs correspond to the C100 cases where $\Lambda_0=100$, which demonstrate the fastest growth of cold gas. The hydro case has no B fields and the plasma $\beta$ panels are thus left blank.}
        \label{fig:cs_viz}
\end{figure*}

\begin{figure*}
     \centering
     
         \includegraphics[width=0.47\textwidth]{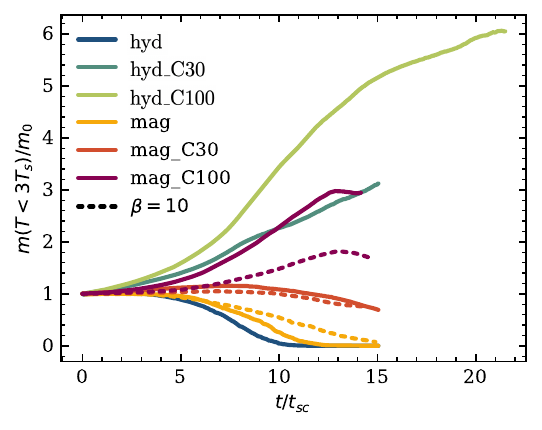}
         \includegraphics[width=0.49\textwidth]{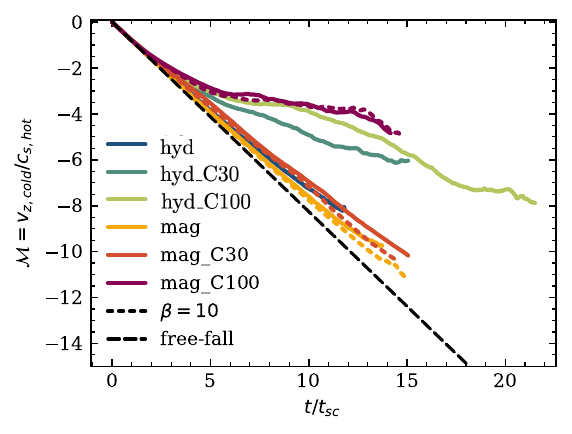}
         \includegraphics[width=0.49\textwidth]{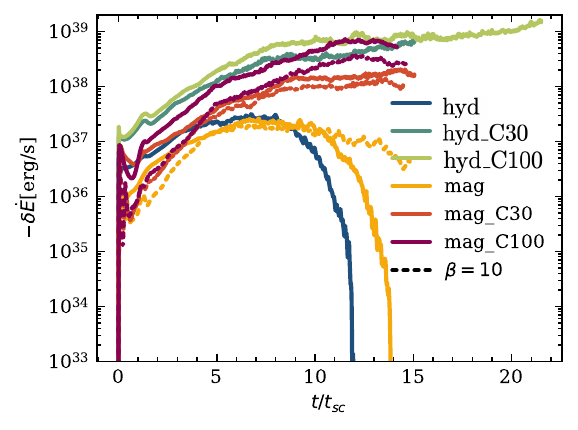}
         \includegraphics[width=0.48\textwidth]{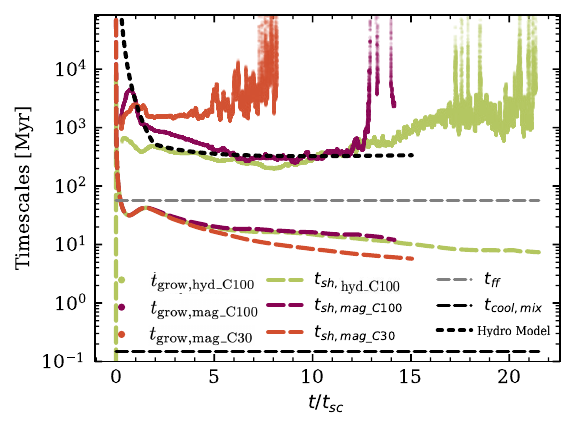}

        \caption{\textit{(Stream, Uniform Background)} Results for the evolution of the cold stream of radius 300pc in a uniform background with (`mag') and without (`hyd') aligned magnetic fields, and varying cooling strengths. (Left Top) The evolution of the cold gas mass as a function of time. (Right Top) Mach number with respect to the hot background gas in the z direction for the cold gas, averaged across the entire stream, along with the free-fall expectation (black dashed line).(Bottom Left) Cooling luminosity evolution. (Bottom Right) Growth and shear timescales comparison overplotted with the $t_{\rm{grow}}$ model (black dotted line) for clouds with $\Lambda_0=100$, albeit with a multiplicative pre-factor that is an order of magnitude higher. Also shown are the free-fall time and cooling time of the mixed gas which is much shorter than any other timescale. 
        }
        \label{fig:nir_grav_mass_vz}
\end{figure*}

Fig. \ref{fig:cs_viz} shows slices of the evolution of a cold stream of $r=300$ pc falling through a uniform background with periodic boundary conditions in $z$, with cooling strength $\Lambda_0=100$, for both the hydrodynamic and MHD ($\beta=100$) case. 
 For streams in a uniform background, as in the cloud case, we only consider aligned fields; we will later consider transverse fields in stratified background. For the adopted parameters, strong radiative cooling enables the stream to not only survive but to grow (we shall re-examine survival criteria in \S\ref{sec:stream-survival}). Fig. \ref{fig:nir_grav_mass_vz} shows how the cold mass, velocity, luminosity, B-fields, and various timescales ($t_{\rm grow}, t_{\rm sh}$) evolve with time. Note that at late times the mixing layer expands to the point where the cold gas starts to leave the box from the sides, which leads to the apparently reduced growth rates (and increased growth times) in the $\Lambda=100$ hydro simulation after $t=15 t_{\rm{sc}}$. This is purely a box size artifact.
 
 In both the slice plots and the plots of time-dependent stream quantities, we see important differences from the cloud case: 
\begin{itemize}
\item{Infall is {\it much} faster, and quickly becomes supersonic (top right of Fig \ref{fig:nir_grav_mass_vz} as compared to Fig \ref{fig:fc_uniform_300pc}; see also shock patterns in slice plots). In many cases, infall is close to ballistic. Even in the strong cooling case ($\Lambda_0=100$) where the stream survives and grows, the stream is still continually accelerating, in contrast to the cloud case, where the cloud quickly reaches a terminal velocity which is much lower than the stream velocities shown here.}
\item{In streams, B-fields amplified by cooling and compression are confined to the stream surface (Fig. \ref{fig:cs_viz}, bottom right panels showing plasma $\beta$), compared to the cloud case, where strong B-fields pervade the cloud (see corresponding panels in Fig. \ref{fig:fc_viz}).}
\item{At a given cooling strength, streams grow much more slowly than clouds of the same radius. For instance, for $\Lambda_0=100, r=300\,$pc, $t_{\rm grow}=400 \, {\rm Myr}$ for streams, but a factor of $\sim 5$ smaller, $t_{\rm grow} \sim 80 \, {\rm Myr}$ for clouds (compare the bottom right panels of Fig \ref{fig:fc_uniform_300pc} and \ref{fig:nir_grav_mass_vz} respectively). } 
\end{itemize}

All of these observations can be understood if mixing is much less efficient in streams than in clouds. In clouds, the interaction between the head of the cloud and the background medium stimulates stronger turbulence and more efficient mixing. Thus, B-fields in the background medium are mixed throughout the cloud. By contrast, mixing in streams only happens at at the surface (and thus, amplified B-fields are only seen at the surface). Less efficient mixing means slower mass growth and a weaker drag force. In this setup, streams therefore accelerate to supersonic velocities under the influence of gravity\footnote{Of course, in a realistic setup, even purely ballistic particles will not accelerate significantly beyond the virial velocity, which is of order the hot gas sound speed. The continual acceleration to supersonic velocities here is an artifact of constant gravity in periodic boundary conditions. Later, we will consider more realistic potentials. However, this artificial setup highlights important differences between mixing in clouds and streams. Note that our analytic expression for stream growth times (equation \ref{eq:tgrow_stream}) is only valid for subsonic and transonic infall, since it does not take the suppression of the KH instability at supersonic velocities into account.}. Once the flow becomes supersonic, mixing and mass growth rates are further reduced, since the mass inflow into the mixing layer stops scaling with Mach number for supersonic flows \citep{YangJi2023}. Note that the critical azimuthal mode number $m$, which is the number of nodes (fixed points) along the circumference of the stream (for azimuthal wavelength $\lambda_{\phi} = 2\pi R_s/m$), for instability increases with Mach number \citep{mandelker16}: 
\begin{equation}
m_{\rm{crit}} = K\left(\left(\frac{\mathcal{M}}{\mathcal{M}_{\rm{crit}}}\right)^2-1\right)^{1/2}
\end{equation}
where $\mathcal{M}_{\rm{crit}} = (1+\delta^{-1/3})^{3/2} \sim 1.3$ with $\delta=100$. Thus, as the Mach number increases, only progressively smaller scale modes are unstable, reducing the efficacy of mixing. This is because smaller wavelengths can only mix at small length scales and therefore it takes much longer for these wavelengths to grow to be on the order of $R_s$ where mixing is maximal. With reduced mixing, the drag force becomes negligible, resulting in quasi-ballistic infall.

\subsection{Uniform Background: Survival Criteria}
\label{sec:stream-survival}

Previous wind tunnel simulations of streams \citep{mandelker20} have found that provided the stream radius exceeds a critical size--similar to the critical radius found for clouds in a wind tunnel--streams can survive and grow. This criterion, where $t_{\rm cool,mix} < t_{\rm sh}$ is analogous to the criterion $t_{\rm cool,mix} < t_{\rm cc}$ for clouds, where the stream destruction time $t_{\rm sh}$ (equation \ref{eqn:tsh}) is substituted for the cloud destruction time $t_{\rm cc}$ (equation \ref{eq:tcc}). We have found that for clouds accelerating under gravity, the survival criterion is modified to $t_{\rm grow} < t_{\rm cc}$, and we might expect the survival criterion for streams accelerating under gravity to be $t_{\rm grow} < t_{\rm sh}$. 

\begin{figure}
    \centering
    \includegraphics[width=0.49\textwidth]{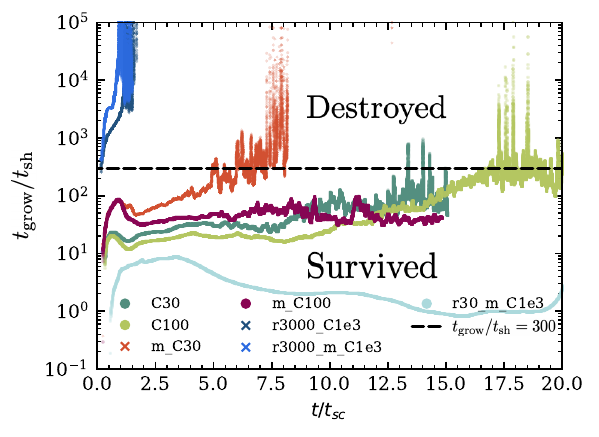}
    \caption{\textit{(Stream, Uniform Background)} Ratio of $t_{\rm{grow}}$ and $t_{\rm{sh}}$ as a function of the time for the streams with various parameters. The dashed line shows our survival criterion for the stream in the uniform background case for the purely hydro and the aligned field $\beta=100$ cases. Crosses in legend denote destroyed streams at the end of the simulation whereas dots denote survived streams.}
    \label{fig:NIR_grav_surv}
\end{figure}

Indeed, we find this to be the case, but with an important modification. Fig \ref{fig:NIR_grav_surv} shows the evolution of the ratio of the growth time to the shear time for various cooling strengths and stream radii. The colors with dots denote streams that survive and grow in mass, while the ones colored with crosses show those that decrease in mass and eventually are destroyed. The dotted line shows the empirically determined survival threshold\footnote{For instance, consider the orange points corresponding to the magnetized stream with a cooling strength $\Lambda_0=30$, where the stream first starts out below but then ends up above the ratio $t_{\rm{grow}}/t_{\rm sh } \sim  300$. Comparing this to the mass evolution of the same case in Fig. \ref{fig:nir_grav_mass_vz}, this turning point corresponds to the place where the stream stops growing and instead starts to lose mass.}, given by $t_{\rm{grow}}/t_{\rm sh } <  300$. Here, the stream growth time is given by equation \ref{eq:tgrow_stream}, which as seen in the black dashed line in Fig. \ref{fig:nir_grav_mass_vz}, accurately models the growth time in simulations (green line). As before, the fact that the C100 simulation (light green points) veers into the destruction regime, is an artifact of stream expansion, so that cold gas starts to leave the box.

Why is the numerical factor in the survival criteria $t_{\rm{grow}}/t_{\rm sh } <  300$ so high? We suspect this is because the expression for $t_{\mathrm{sh}}$ (equation \ref{eqn:tsh}) becomes inaccurate for highly supersonic speeds; it overestimates mixing and significantly underestimates the destruction time. The obvious disparity can be seen when $\mathcal{M}>1$, $t_{\mathrm{sh}}\propto\frac{1}{\mathcal{M}}$ which is incorrect; the stream needs at least one sound crossing time to be disrupted. We thus turn to the disruption criteria for body modes instead fo surface modes \citep{mandelker19}. The shear timescale $t_{\rm sh}$ is accurate for at most transonic stream velocities where surface modes are responsible for disruption. In the superosonic regime, the disruption is dominated by body modes which has a characteristic timescale 
\begin{equation}
    t_{\rm dis} = t_{\rm sc} \left(1+0.5\ln{\left(\frac{R_s}{H_0}\right)}\right)
\end{equation}
where $H_0$ is the initial amplitude of perturbations to the shape of the stream. After an initial instability growth phase, we can assume the shape perturbation is 1 cell in amplitude, and the stream is resolved by 32 cells. This means $t_{\rm dis}\sim t_{\rm sc}(1+0.5\ln(32))\sim 2.73 t_{\rm sc}$. Using this instead of $t_{\rm sh}$ and assuming a fairly constant shear timescale value from simulations $t_{\rm sh}\sim 10 \, \rm {Myr}$, we get the new criteria for survival
\begin{equation}
    t_{\rm grow}\lesssim25 t_{\rm dis}
\end{equation}

In principle, we should find a new expression for $t_{\rm sh}$ in the supersonic regime in a series of adiabatic simulations with supersonic shear velocities. Such simulations would be relevant also for the survival of supersonic jets. However, for more realistic stratified setups, streams do not accelerate to supersonic velocities. As previously noted, even ballistic particles cannot accelerate to super-virial velocities in a realistic potential. We therefore do not further consider the behavior of highly supersonic streams, which we regard as beyond the scope of this paper.

\subsection{Stratified background}
\begin{figure*}
         \includegraphics[width=0.48\textwidth]{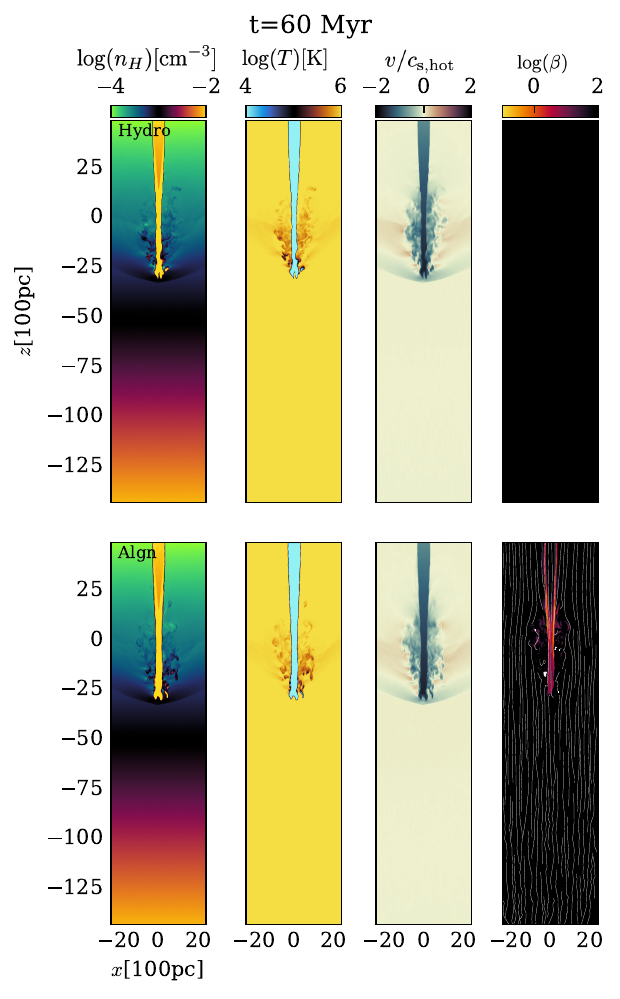}
         \includegraphics[width=0.48\textwidth]{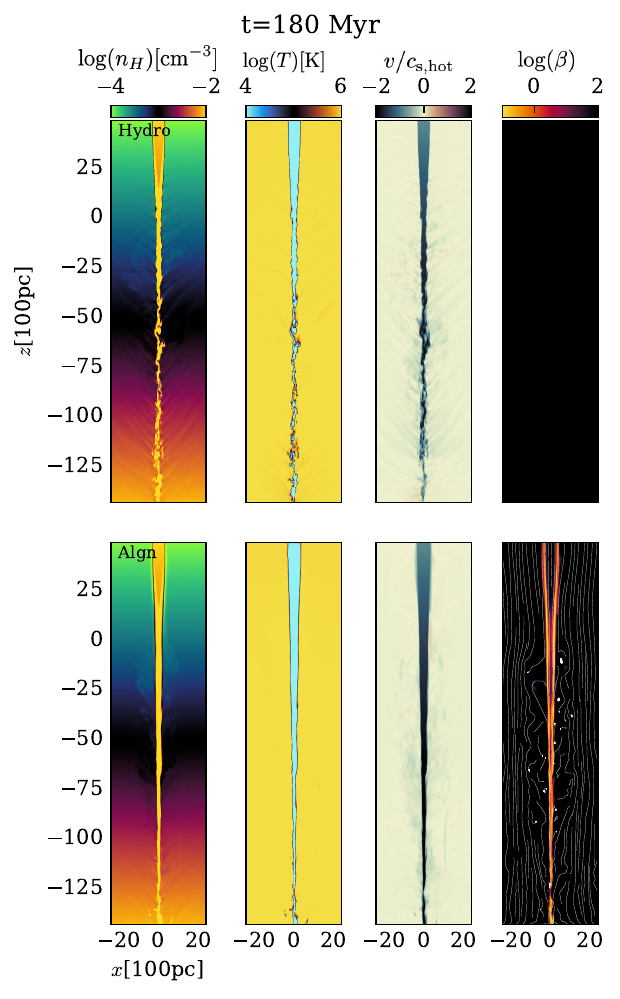}
        \caption{\textit{(Stream, Stratified Background)} Evolution snapshots of an infalling stream of radius $r=300$ pc falling under gravity in a stratified background showing slices of density (first column), gas temperature (second column), velocity in the z direction in units of the hot gas sound speed (third column) and magnetic fields along with field lines (fourth column) for the purely hydro case (top row) and aligned B fields case (bottom row) at $t=60$ Myr (left panel) and $t=180$ Myr (right panel). The hydro case has no B fields and the B field panel is thus left blank.}
        \label{fig:cs_strat_viz}
\end{figure*}

\begin{figure*}     
         \includegraphics[width=0.49\textwidth]{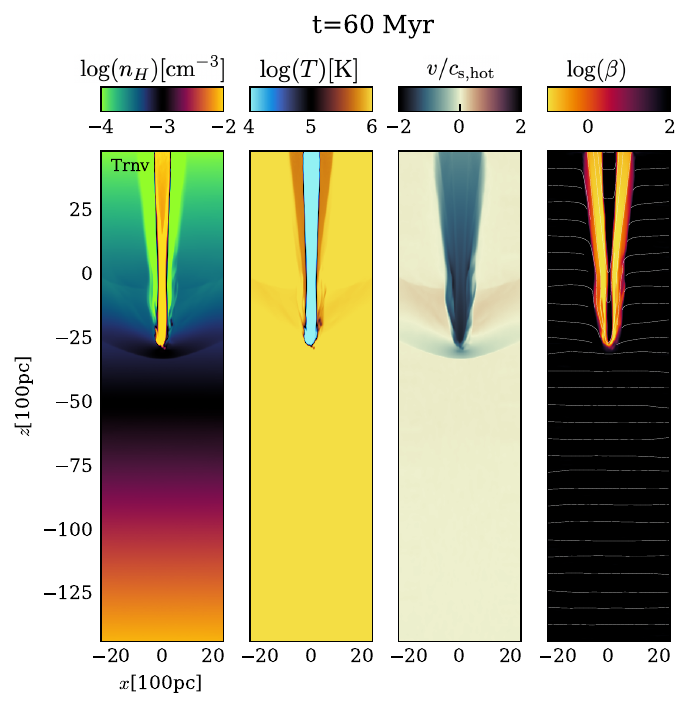}
         \label{fig:cs_strat_viz1_trnv}
         \includegraphics[width=0.49\textwidth]{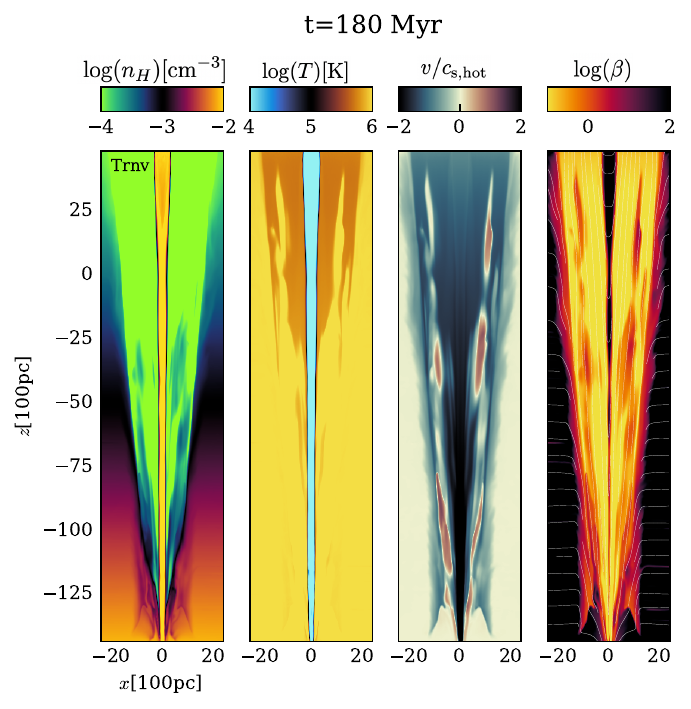}
        \caption{Same as Fig. \ref{fig:cs_strat_viz} but for transverse fields}
        \label{fig:cs_strat_viz_trnv}
\end{figure*}

We now present results for a stream falling through a stratified background. 
Similar to the cloud, we consider a constant gravity isothermal background. However, because the stream extends throughout the atmosphere, we cannot use the same shortcut we used for the cloud (a frame-tracking algorithm where the background hot gas increases in density as the cloud falls). Instead, we must simulate the entire extent of the stream. We create a long box which is $\sim 7$ density scale heights in length, where the hot gas is in hydrostatic equilibrium. A stream of cold gas is injected at the top boundary with velocity $0.8c_{\rm{s,hot}}$ (we are not sensitive to the exact injection velocity). This gas falls and eventually leaves the bottom of the box, establishing a time-steady profile. A similar, more realistic setup was adopted by \cite{Aung2024} (see \S\ref{sec:realistic}) which had the stream roughly fall through $\sim 6.5$ scale heights through the virial radius of the galaxy.

Fig. \ref{fig:cs_strat_viz} shows the slices from the evolution of the hydrodynamic and aligned B-field cases ($\beta=100$) at two different points in time. 
Compared to the uniform background case, a stronger velocity gradient along the streamwise direction is apparent. The hydrodynamic case also shows sinusoidal ripples (more on this shortly), while the aligned MHD case is comparatively smooth. In the latter, the interface between hot and cold gas is strongly magnetized (low $\beta$), as a result of compressional amplification of hot gas which has cooled. Fig. \ref{fig:cs_strat_viz_trnv} shows the snapshots from the evolution of the transverse field case ($\beta=100$). The horizontal extent of the stream is much larger in this case. Magnetic draping of initially horizontal fields creates a broad interface of magnetically supported gas, most of it still hot. The inner cold core of infalling gas is still relatively pristine and unaffected. 

Fig. \ref{fig:cs_strat_plots} shows the evolution of the falling cold stream with $z$, averaged across 55 Myr (corresponding to 90 simulation snapshots) in the steady-state regime. 
Since the stream falls from positive to negative z, the plots should be read from right to left. The top two plots depict the mass per unit length: 
\begin{equation}
    m = \left\langle\frac{dM}{dz}\right\rangle_{x,y} = \langle \rho A \rangle_{x,y}
\end{equation}
and the mass flux: 
\begin{equation}
\langle m v_z \rangle_{x,y} =  \langle \rho A v_z \rangle_{x,y} 
\end{equation}
where $A$ is the cross-sectional area, and $\langle\cdots\rangle_{x,y}$ represents an area average over the $x$ and $y$ directions over the quantity inside the brackets for all gas satisfying $T<3\times 10^4$ K. As the data is quite noisy, and exhibits oscillations (see below for further discussion), we smooth it in two ways: we apply a smoothing B-spline (basis-spline) interpolation \citep{DEBOOR197250} to the curves as a function of height ($z$), and also implement a running mean over 10-25 points for each of the curves in $z$ to provide extra smoothing over the remaining noise. This is necessary to smooth over bumps in the mass evolution curves where $\dot{m}$ that would result in unphysical infinite growth times. Finally, since we now have time-averaged $m(z),\dot{m}(z)$ rather than $m(t),\dot{m}(t)$, we infer the mass growth time in the steady-state regime as: 
\begin{equation}
    t_{\rm{grow}}(z) = \left\langle\frac{m}{\dot{m}}\right\rangle_{x,y} = \left\langle\frac{m}{\frac{dm}{dz}\dot{z}}\right\rangle_{x,y} = \left\langle\frac{m}{\frac{dm}{dz}v_z}\right\rangle_{x,y}
\end{equation}

Let us first focus on the hydrodynamic results (solid blue) in Fig. \ref{fig:cs_strat_plots}. Perhaps the two most important questions with regard to stream-halo interactions are: (i) does cold mass growth due to mixing and cooling increase the mass flux to the central galaxy? (ii) Is the luminosity due to this process observable? In this setup, the mass flux $\langle m v_{\rm z} \rangle$ does increase by a factor of a few (top right plot). Since the stream is continually falling towards denser gas, the mass growth time grows shorter, and the cooling luminosity increases, as the gas falls inwards (bottom left and middle right plots). Once mass growth and hence accretion braking become efficient at $z \sim -7.5$kpc, the velocity decreases sharply. Since $t_{\rm grow}$ continually decreases, 
accretion braking grows more and more effective, and the stream eventually decelerates towards lower velocities. 
The mass per unit length, mass flux, and infall velocity are all in reasonable agreement with the analytic model (dotted black line; the coupled equations \ref{eqn:stream_eq1} and \ref{eqn:stream_eq2} along with scalings from \ref{eq:tgrow_stream}). 

How do magnetic fields affect these outcomes? We can see that aligned magnetic fields (green lines) strongly inhibit mixing and mass growth, so that $t_{\rm grow}$ is larger than the hydrodynamic case by an order of magnitude (bottom left panel of Fig. \ref{fig:cs_strat_plots}). Thus, the mass flux $\langle m v_{\rm z} \rangle \approx$const, i.e. cold gas mass is conserved, with only a slight upturn (for the $\beta = 100$ case) near the bottom of the box (top right panel). Note that the cooling luminosity is already strongly suppressed for the $\beta =100$ (middle right panel), and increasing the field strength (to $\beta=10$, dashed green lines) only provides modest further suppression. This is because gas in the interface between hot and cold gas already has low $\beta$ due to compressional amplification (Fig \ref{fig:cs_strat_viz}), and lowering the background $\beta$ does not significantly change $\beta$ at the interface 
The lack of significant accretion braking means that the stream accelerates at close to the ballistic rate (green, grey lines in the central left panel). This increase in velocity means that since mass flux is conserved $\langle m v_{\rm z} \rangle \approx$const, the mass per unit length $\langle m \rangle = \langle \rho A \rangle$ decreases as the stream falls (top left panel).

\begin{figure}
    \centering
    \includegraphics[width=0.99\linewidth]{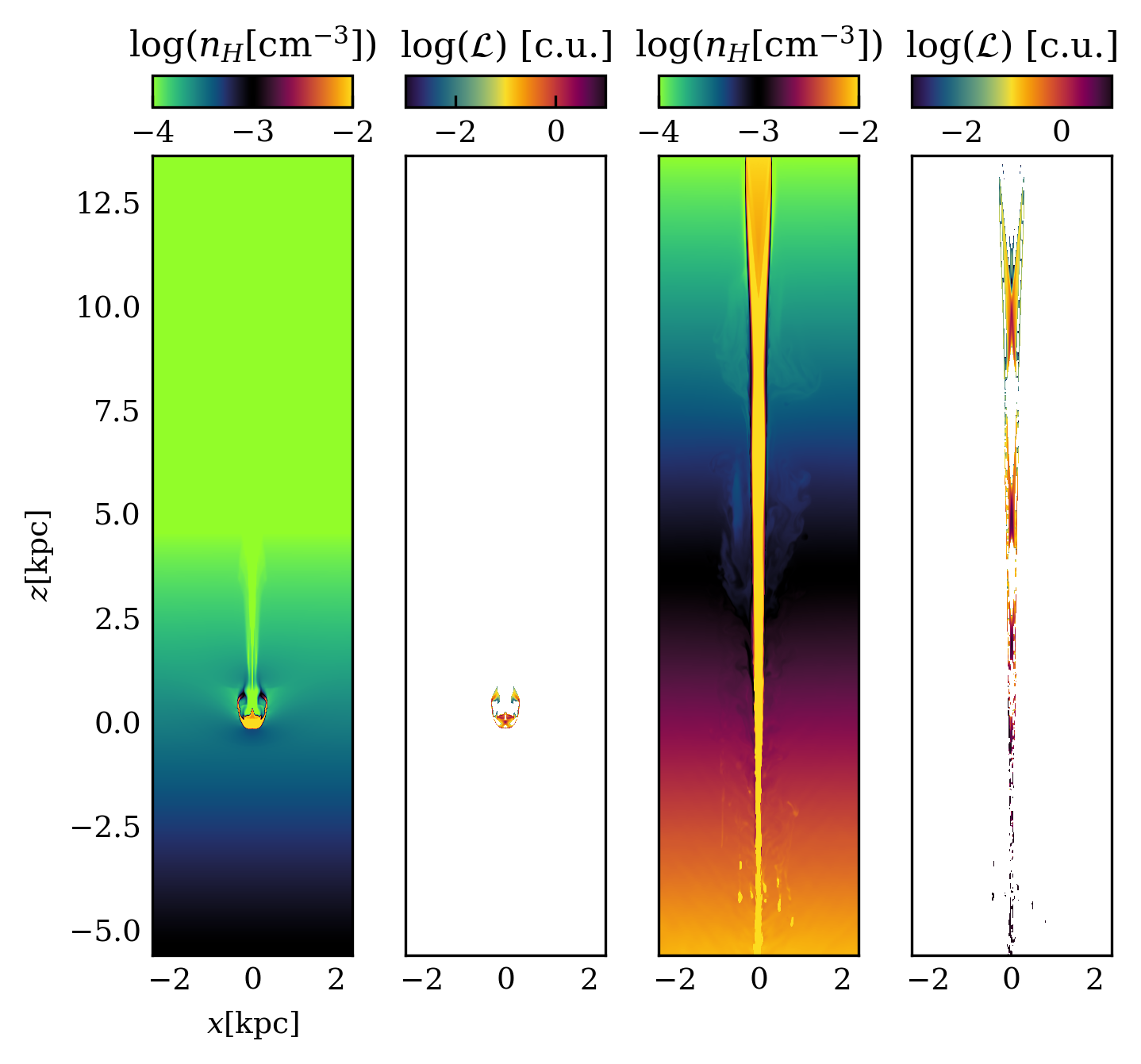}
    \caption{The locations along infalling clouds and streams where cooling occurs for aligned magnetic fields initialized with $\beta=100$ in an isothermal hydrostatic density-stratified environment. (Left Panels) The number density and luminosity of the infalling cold cloud at early times, highlighting the regions where cooling and mixing are taking place. Here cooling occurs at the head and sides of the cloud. (Right Panels) Same as the left two panels but for the stream. Here cooling mostly occurs in the bottom half of the box, and is sporadic in the top half.}
    \label{fig:fc_cs_algn}
\end{figure}

\begin{figure*}   
         \includegraphics[width=0.48\textwidth]{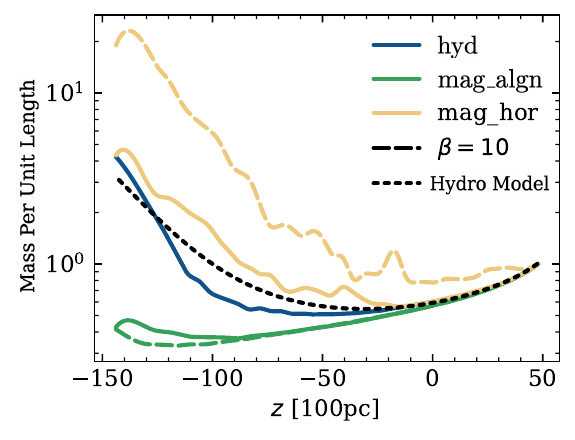}
         \includegraphics[width=0.48\textwidth]{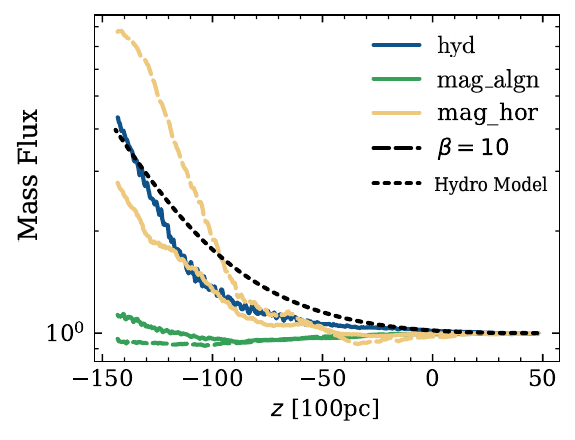}
         \includegraphics[width=0.48\textwidth]{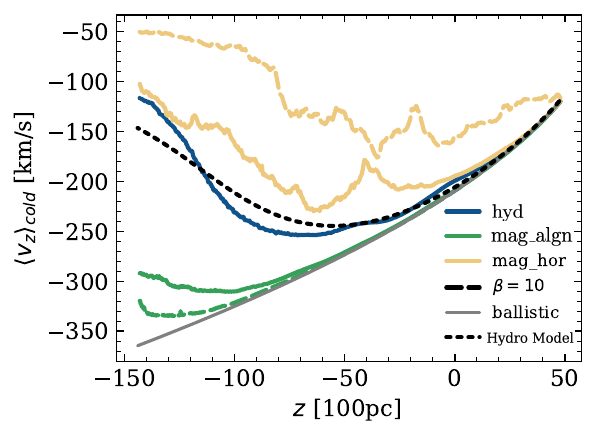}
         \includegraphics[width=0.48\textwidth]{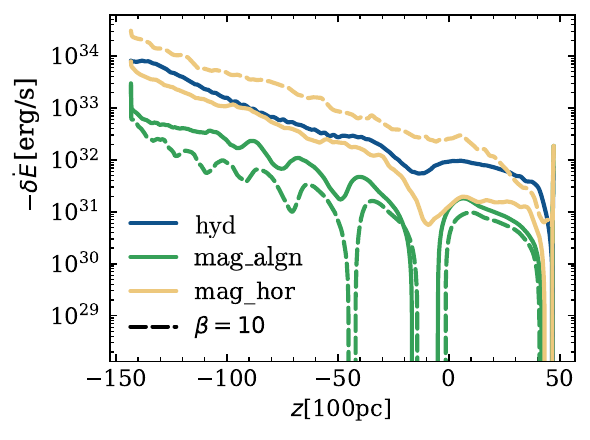}
         \includegraphics[width=0.48\textwidth]{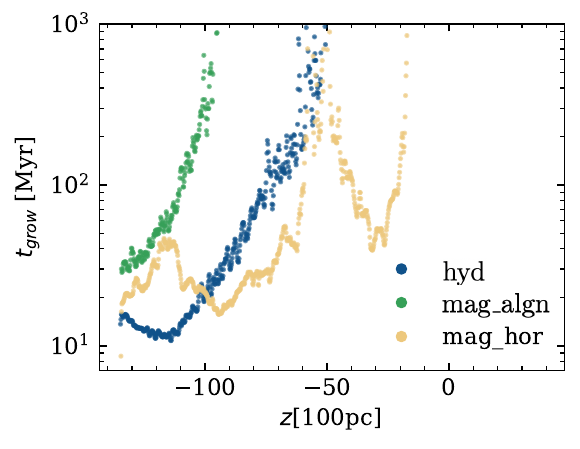}
         \label{fig:cs_t_grow}
         \includegraphics[width=0.48\textwidth]{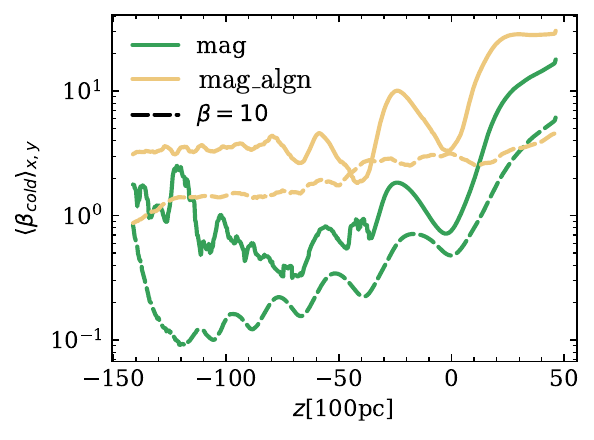}
        \caption{\textit{(Stream, Stratified Background)} 
        Results for the evolution of the cold stream of radius 300 pc with an injection velocity of $0.8c_{\rm{s,hot}}$ falling through a stratified background. (Top Left) The evolution of the cold gas mass as a function of the vertical height (z-direction). (Top Right) Evolution of the cold mass flux with z. (Middle Left) Velocity in the z direction for the cold gas as a function of z, (x and y) averaged across the entire stream radius. (Middle Right) Cooling luminosity evolution vs $z$. (Bottom Left) Growth timescale across $z$. (Bottom Right) Evolution of the cold gas plasma parameter $\beta$ with $z$. The black dashed lines show the hydrodynamical infalling stream model (equations \ref{eqn:stream_eq1}, \ref{eqn:stream_eq2} with equation \ref{eq:tgrow_stream} scalings) evaluated for the isothermal stratified background. These results are averaged over $\sim 50$ Myr beginning from $300$ Myr, by which time the stream head has crossed the bottom boundary. Due to compressional amplification of B-fields that has been occurring, even though $\beta_0=100$ everywhere initially, these curves at late time begin (at the top of the box) at lower $\beta$ than at the initial condition.}
        \label{fig:cs_strat_plots}
\end{figure*}

Why are aligned magnetic fields more effective at preventing cold gas accretion onto streams but not clouds? Recall that in the cloud case, aligned magnetic fields do not significantly change the dynamics or growth. However, for streams even weak ($\beta_0=100$) aligned fields reduce mass growth and cause streams to fall at ballistic velocities. The difference arises from the distinct regions of cooling and mixing in clouds versus streams, as shown in Fig. \ref{fig:fc_cs_algn}. During the early evolution of clouds, mixing and cooling occurs both at the sides and head, where fresh hot background gas is continuously encountered. In contrast, streams lack a distinct "head" region, which means that the primary site for cooling and mixing is confined to the stream’s interface with the background gas. Because aligned magnetic fields suppress mixing at the interface, streams experience reduced mass growth and sustain higher velocities, leading to near-ballistic infall. 

The results for the transverse field case (beige lines in \ref{fig:cs_strat_plots}) are even more interesting. In the cloud case, magnetic draping inhibits mixing and mass growth, but magnetic drag (rather than accretion drag) slows the cloud. For the infalling streams, magnetic drag also plays a significant role in slowing down the stream, 
which falls more slowly than the hydrodynamic case; the effect becomes more pronounced for higher background magnetic fields (witness the strong deceleration in the $\beta=10$ case). However, the mass flux and the cooling luminosity are {\it not} suppressed for the $\beta=100$ case, and in fact they are significantly {\it enhanced} relative to the hydrodynamic case for $\beta=10$. 
The physics is similar to the case for clouds (\S\ref{subsec:clouds_transverse}). Cold gas falls significantly more slowly in the transverse field case, particularly when the field is strong. A Lagrangian fluid element at a given $z$ in the transverse field case is much `older' than its hydrodynamic counterpart, and has had more time to mix. The lower infall velocity means that for a given mass flux $\langle m v \rangle$, the mass per unit length $\langle m \rangle$ is larger and there is more gas available for mixing. This increase in $m$ can compensate for an increase in $t_{\rm grow}$ due to suppression of mixing, so that $\dot{m} \approx m/t_{\rm grow}$ can be similar or even larger than the hydrodynamic case. For instance, consider the $\beta=10$ transverse field case (dashed beige lines), where these effects are strongest. As gas falls from $z=5$kpc to $z=-9$kpc, the mass flux is comparable to or less than the hydrodynamic mass flux (blue lines). However, because the infall velocity is much lower (middle left plot), the mass per unit length $\langle m \rangle$ is much higher (upper left plot), and the corresponding mass growth rate $\dot{m} \propto \dot{E}$ at a given $z$ is higher than the hydrodynamic case (middle right panel). For $z < -9$kpc, the mass per unit length is so much larger, that even the mass flux $\langle m v \rangle$ is larger than the hydrodynamic case. One might worry that magnetic tension effects are sensitive to box width. We have confirmed that we reproduce the same results if we double the box width. In summary, {\it an increase in the infall time can increase the total mass flux to the central galaxy, since the stream has more time to grow.} 

Even after the smoothing procedures we have applied, there are conspicuous oscillations visible in the plots (particularly in the cooling luminosity), which are also visible in the stream morphology. These arise when the flow becomes supersonic with respect to the background medium. The stream takes a finite time to react to the change in confining pressure as it falls in a stratified medium. Internal, reflecting oscillations develop which create a series of over and under dense/pressured regions in the stream, known as `shock diamonds'.  They are also seen in supersonic jet flows from nozzles and have been extensively studied in terrestrial applications. Since the aligned magnetic field cases in Fig. \ref{fig:cs_strat_plots} are the \textit{most} supersonic, the shock diamonds are most evident in this case. They do not affect the secular evolution of the stream, and mostly create periodic variations in the stream profile which we suppress with our smoothing procedures. 

\subsection{Realistic Background}
\label{sec:realistic}

A $g=$ const, isothermal density profile might be realistic near the disk, but it is not realistic through a very large number of scale heights. In a realistic halo, the gravitational force is a function of radius, and there is also a well defined central density (which is the well defined center to the potential well) and a limited number of density scale heights that the stream falls through. We address this by using a more realistic density profile exactly like the one used by \citet{Aung2024}. We use an NFW dark matter density profile appropriate for a $z\sim 2$, $M=10^{12}M_{\odot}$ halo with a concentration parameter $c=10$ with a virial radius of $R_{\rm vir}=100$kpc and a virial velocity of $v_{\rm vir} = 200 \, {\rm km \, s^{-1}}$. We assume the $T\sim T_{\rm vir} \sim 10^6{\rm K}$ gas to be in hydrostatic equilibrium, using $n\sim10^{-4} \rm{cm}^{-3}$ at $R_{\rm{vir}}$ with the bottom (inner) boundary conditions set by the \cite{Aung2024} profile. This yields a slightly increasing temperature profile towards the bottom and a density that has increased by $\sim 2.5$ orders of magnitude. Even though we are simulating a halo, we retain Cartesian geometry for simplicity, mapping the radial coordinate to the height coordinate. Our box of $[0.5R_{\rm{vir}}, 0.5R_{\rm{vir}}, R_{\rm{vir}}]$ is resolved by $256\times 256\times 512$ cells. Similar to the stratified isothermal halo case, the stream is then injected from the top at $1.1 R_{\rm{vir}}$ and we follow the profile of the stream as it travels to the disk to $0.1 R_{\rm{vir}}$. While our default case considers solar metallicity, we also consider a more a realistic $Z=0.1 Z_{\odot}$ metallicity. For the MHD simulations, we adopt $\beta=100$.

\begin{figure*}
         \includegraphics[width=0.48\textwidth]{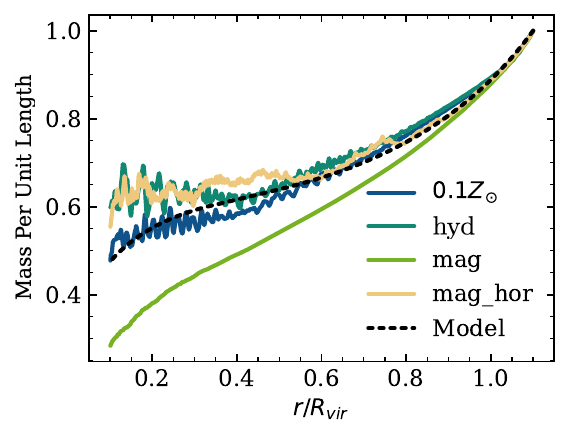}
         \includegraphics[width=0.48\textwidth]{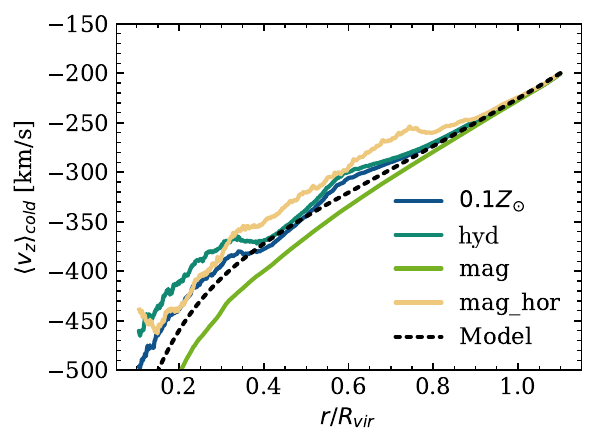}
         \includegraphics[width=0.48\textwidth]{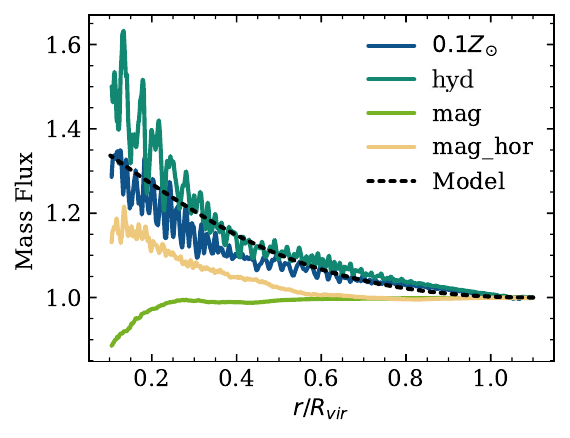}
         \includegraphics[width=0.48\textwidth]{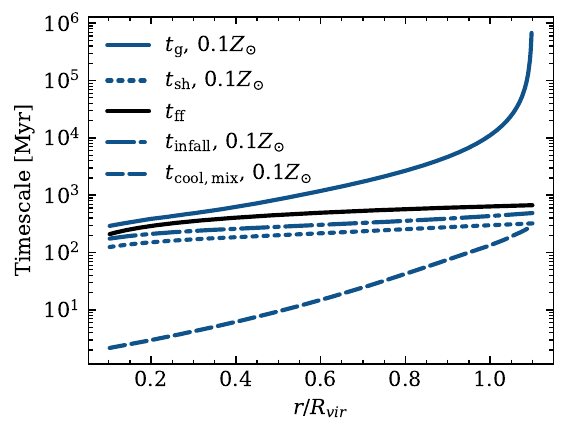}
        \caption{\textit{(Stream, Realistic Background)} Results for the evolution of the cold stream of radius 3 kpc with an injection velocity of $V_{\rm{vir}} = $ 200 km/s falling through a realistic NFW background along with the model prediction. (Top Left) The evolution of the cold gas line mass fraction as a function of the vertical height (z-direction).(Top Left) Velocity in the z direction for the cold gas as a function of z, averaged across the entire stream. (Bottom Left) Evolution of the cold line mass flux fraction with z.(Bottom Right) Comparison of the growth time, shear time, free-fall time, infall time and the mixed gas cooling time for the $0.1Z_{\odot}$ simulation. 
        }
        \label{fig:cs_real_plots}
\end{figure*}

Fig. \ref{fig:cs_real_plots} shows the evolution of the stream for all the different models. For this more realistic case, the growth time is sufficiently long compared to the free-fall time that there is comparatively little mass growth -- the mass flux increases by only $\sim 30\%$ in the hydrodynamic case. As a result, the stream falls quasi-ballistically. The acceleration means that for conserved mass flux, the mass per unit length drops. Magnetic fields in the $\beta=100$ case make little difference. For aligned fields, the mass flux is roughly constant $\langle m v \rangle =$const, due to suppression of mixing, so that the stream's trajectory is fully ballistic. For $\beta=100$ transverse field, the mass flux is intermediate between the hydrodynamic and aligned field case. 
 For the $\beta=10$ case (not shown) mass flux does not increase substantially -- even though there is much more braking, and thus more mass accretion on to the stream, both of these effects cancel out to yield mass flux similar to the purely hydrodynamic case. Finally, we see that the reduction in cooling efficiency for $Z=0.1 Z_{\odot}$ only has a minor effect. 

The hydrodynamic result seems to be in disagreement with \cite{Aung2024}, who find that the mass flux increases by $\sim 2$ and the mass per unit length increases by $\sim 1.3$. This discrepancy is because of the times chosen to evaluate the profiles. \cite{Aung2024} evaluate profiles after one box crossing time, but we evaluate them after 5-6 box crossing times. We confirm that the line mass and mass flux profiles are generally consistent with \cite{Aung2024} if we evaluate them at early times.

\begin{figure}
    \centering
    \includegraphics[width=\linewidth]{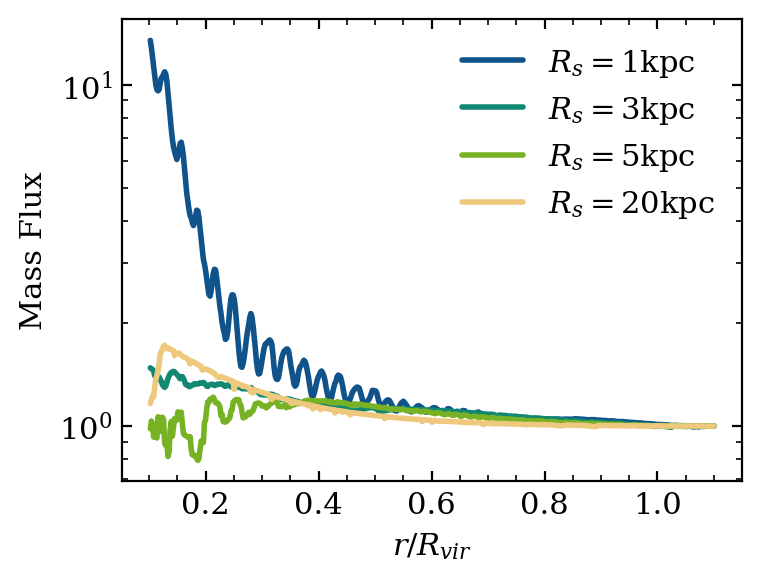}
    \caption{The variation of mass flux with height of hydrodynamical infalling streams with four different initial radii in a realistic background. These radii are chosen to be typical values of stream radii in $10^{12}M_{\odot}$ halos as predicted by analytic theory \citep{mandelker20-blobs}. No stream with these radii is destroyed. Instead each simulation has more growth as compared to the adiabatic non-radiative case, which would simply be a constant line at unity (since mass flux must be conserved).}
    \label{fig:cs_real_sizes}
\end{figure}

Infalling cold streams are also very robust to destruction. Fig. \ref{fig:cs_real_sizes} shows, that the mass flux of hydrodynamical streams falling through this realistic background always increases for reasonable choices of initial stream radii, based on analytic models for a $10^{12}M_{\odot}$ halo at $z\sim 2$ \citep{mandelker20-blobs}. The reason for this is the long stream disruption time as compared to the infall time. As shown previously, the stream disruption time for body modes for a 3kpc stream would be $t_{\rm dis}\sim 3t_{\rm sc}\sim 6R_{s}/c_{\rm s, cold}\sim 10^3 \, \rm {Myr}$, while the infall time, assuming $v\sim 300 \, \rm km\,  s^{-1}$, is $t_{\rm infall}\sim R_{\rm vir}/v_{s}\sim 300\,\rm{Myr}$. In fact, $t_{\rm infall}$ becomes shorter and shorter as the stream gets closer to the inner radii. Since $t_{\rm infall}/t_{\rm dis}\leq1$ typically, for a realistic range of parameters, such streams are difficult to disrupt and can easily survive the jouney to the inner halo of the galaxy. 

\section{Other Applications}
\label{sec:discussion}
Most of this work has focused on the survival of infalling clouds and streams in galactic halos, which are thought to be critical for supplying gas for star formation. Here, we briefly discuss potential applications to two somewhat more specialized phenomena: compact high velocity clouds (\S \ref{sec:compact}) and coronal rain (\S\ref{sec:rain}).

\subsection{Survival of Compact High Velocity Clouds}
\label{sec:compact}
\citet{tan23-gravity} predict, based on their survival criterion, that infalling clouds should survive if they are within 10 kpc of the galactic disk. This prediction agrees well with observations of high-velocity clouds (HVCs) whose distances can be constrained through stellar bracketing (i.e., by observing the occurrence, or lack thereof, of absorption lines in stars with known distances). They further argue that at greater distances, larger clouds may still survive if they can reach this ''growth zone" without being completely disrupted. Such clouds, upon fragmentation, would likely appear as a complex of smaller clouds. 

Interestingly, a subset of HVC observations includes small, distant clouds known as compact high-velocity clouds (CHVCs). These CHVCs have small angular sizes and tend to be located at distances on the order of a few hundred kpc \citep{braun99,putman02}. It is argued that the population of CHVCs is distinct from the general HVC population, with different overall properties. Their median velocities are suggestive of infall toward the center of the Local Group. The existence of CHVCs is puzzling, since they should be disrupted in just a few hundred Myr by hydrodynamical instabilities. At such distances, cooling should also be inefficient.

We have shown that fields with weak background transverse magnetic fields both enhance cloud survival and do not inhibit growth at later stages of the infall. This can potentially explain both observations of smaller, distant clouds as well as nearer HVCs. Distinguishing between hydrodynamic and MHD models would require more direct comparisons of less idealized models with observations. For example, observations of magnetic field strengths could determine whether accretion or magnetic drag is more important. Ultimately, this would affect the amount of cold gas which eventually reaches the disk.

High-resolution cosmological galaxy formation simulations, such as TNG50, find that thermal instability in the CGM can generate cold clouds that are magnetically dominated \citep{nelson20}. These clouds are found to be long-lived, and show a range of magnetic field draping and orientations with respect to the cloud morphology \citep{ramesh24}. Observationally, however, magnetic field strengths in HVCs remain poorly constrained. Current measurements often rely on identifying coherent structures in rotation measures, which are primarily sensitive to larger-scale magnetic coherence \citep{mcclure10}. This limitation underscores the need for more detailed observations to fully understand the role of magnetic fields in the survival and evolution of these clouds.

\subsection{Survival of Coronal Rain}
\label{sec:rain}
Numerous observations of the solar corona (atmosphere) have shown that it is composed of a similar environment as the CGM, i.e. it hosts $10^6$ K hot background gas in addition to $10^4$ K cold condensations that occur in highly magnetized coronal loops, termed coronal rain \citep{antolin2010, antolin2012, sahin2023}. This coronal rain has been proposed to be a marker of the long-standing solar coronal heating problem \citep{antolin2010} as well as being an important tracer of the magnetic field and topology of the corona \citep{kriginsky2021}. Coronal rain is naturally a direct analogue to infalling clouds in the CGM, albeit with strong magnetic fields ($\beta\sim 0.01$), and has been known to fall at speeds far below free-fall \citep{sahin2023}, with terminal velocities which depend on overdensity. The accretion braking mechanism studied here under the presence of aligned fields is thus one potential way to slow down these clouds to a terminal velocity in this environment, a process that cannot be studied in simulations which model rain as overdense blobs where radiative cooling is ignored \citep{martinez2020}. Moreover, the formation and dynamics of coronal rain is tracked in real time -- on timescales of minutes to days–
in both emission and absorption in a broad range of spectral lines. These include Ly$\alpha$, H$\alpha$, as well as lines associated with transition temperatures such as HeII 304 \AA, OV and OVI, SiIV and SiVII. The observed emission can be translated into a mass accretion rate and accretion drag force, to check for consistency with the observed dynamics. 

\section{Conclusions }
\label{sec:conclusions}
In this work we analyze the impact of magnetic fields on radiatively cooling, infalling cold clouds and streams in the CGM. We simulate cold clouds and streams falling through uniform
and stratified hot backgrounds with and without magnetic fields with aligned and orthogonal orientations. Previous work which studied the effect of magnetic fields on radiatively cooling clouds and streams did so in wind tunnel setups. These have qualitatively different behavior and survival criteria. The drag force caused by mass entrainment in the infalling case is insufficient to overcome the gravitational acceleration, and thus there is always a net acceleration down the potential well, leading to continual high levels of shear. Our work is an extension of previous hydrodynamic simulations of infalling clouds \citep{tan23-gravity} and streams \citep{Aung2024}. Our conclusions are as follows: 
\begin{itemize}

  \item{For infalling clouds, the inclusion of aligned magnetic fields negligibly impacts mass growth, cold cloud velocity, cooling emission and growth rates compared to the hydrodynamic case, for weak background magnetic fields ($\beta=P_{\rm thermal}/P_{\rm mag} \sim 10-100$) expected in the CGM. Clouds grow in mass, and experience `accretion drag' due to the accretion and incorporation of cooling hot gas, which is initially static and has zero momentum. This gives a survival criterion $t_{\rm grow} \lsim 4t_{\rm cc}$, which differs from the wind tunnel case, and a terminal velocity $v_{\rm T} \sim g t_{\rm grow}$ \citep{tan23-gravity}. Here, $t_{\rm grow}, t_{\rm cc}$ are the cloud growth and destruction times, which have analytic expressions. 
  
  \item For stronger fields, $\beta \sim 1$, mass growth is suppressed (growth times are a factor of $\sim 2$ long), and cold gas survival in marginal cases is somewhat enhanced over hydrodynamic simulations, due to reduced mixing. The weak effects of magnetic fields is consistent with previous wind-tunnel simulations \citep{gronke20-cloud}. The scaling of mass growth rates with turbulent velocity $\dot{m}\propto \sigma_{\rm v, turb}^{3/4}$ inferred in hydrodynamic turbulent mixing layer simulations still holds even with the inclusion of magnetic fields, both weak and strong. Any reduction in mass growth is caused by the suppression of turbulence by magnetic tension.}

   \item By contrast, even weak transverse magnetic fields ($\beta \sim 10-100$) have a strong impact on infalling clouds. The reason is magnetic draping: magnetic fields are swept up around the infalling cloud and strongly compressionally amplified. The usual expectation for draped magnetic fields is equipartition with hydrodynamic ram pressure. In wind tunnel simulations, draped fields can extend cloud lifetimes, but -- similar to the aligned fields case -- do not dramatically change outcomes or growth rates. However, for infalling clouds, the plasma beta in the drape ($\beta \ll 1$) is much smaller than in the wind tunnel case. This is due to `velocity overshoot': clouds falling under gravity accelerate to high velocities $v_{\rm max}$ before drag forces kick in to decelerate the cloud to a significantly lower terminal velocities $v_{\rm T,accrete}$. Strong magnetic fields arise due to the high ram pressure $\rho v^2_{\rm max}$ at peak infall velocity, and the plasma $\beta$ in the drape remains small after deceleration.  

   \item The low plasma $\beta$ in the drape has several consequences. Mixing is reduced by magnetic tension, so that mass accretion rates $\dot{m}$ are lower and growth times $t_{\rm grow}$ are higher. Even more imporantly, the stronger fields imply that magnetic drag forces $F_{\rm drag} \sim \rho v_{\rm A}^2 A_{\rm cloud}$ are much stronger than accretion drag. This leads to a new terminal velocity $v_{\rm T,MHD} \sim (\delta R g - v_{\rm A}^2)^{1/2}$ (where $\delta$ is the cloud overdensity) which can be much smaller than the terminal velocity due to accretion drag $v_{\rm T,accrete} \sim g t_{\rm grow}$. The low terminal velocity extends the infall time, leading to the surprising result that {\it even for $\beta \sim 10-100$, transverse fields enhance overall mass growth compared to the hydrodynamic case} (middle and right panels of Fig. \ref{fig:fc_strat_mag_hor_vs_z}), with a stronger effect for stronger fields. Thus, transverse magnetic fields can boost the overall cold gas inflow into galaxies. 

   \item Infalling cold streams show qualitatively similar behavior. However, both mass growth and accretion drag are significantly weaker. An important difference is that clouds have a `head' where a good deal of mixing and mass growth take place, whereas steady-state streams only allow for lateral mixing. Thus, for instance, unlike clouds,  cold streams falling in a uniform background fall almost completely ballistically, accelerating to highly ($\mathcal{M} \gg 1$) supersonic velocities. Even more importantly for realistic scenarios, aligned magnetic fields strongly suppress mass growth and accretion braking in streams, even at high $\beta$ ($\beta\sim100$) and in a stratified background\footnote{A technical point: in our simulations, for aligned magnetic fields the background $\beta$ increases as the stream falls, since $P_{\rm gas}$ increases but $P_{\rm B}$ has to remain constant in Cartesian geometry for the magnetic field to be divergence free.}. 
   
   \item At the same time, the strong fields in the mixing region (which are amplified by flux freezing as magnetized hot gas cools) protect the stream against mixing and destruction, producing a roughly constant mass flux even as the stream accelerates to high velocities. This is in contrast to clouds, where an identical simulation setup shows little difference in growth and accretion drag between the hydrodynamic and aligned B-field cases. Thus, streams closely align with the results of MHD turbulent mixing layer simulations, which show strong suppression of mixing and mass growth due to magnetic tension \citep{ji19,zhao23}, whereas in clouds the additional dynamics associated with the cloud head changes matters \footnote{For this reason, streams which are already existing in the simulation box give different results from streams which are `dropped' in a gravitational field and initially have a head. For instance, `dropped' streams have strong magnetic field throughout the stream, whereas streams run with periodic boundary conditions only have significant B-fields in the mixing layers on the side.}. 
   
   \item Transverse magnetic fields show the closest similarity to the cloud case, and do slow streams via magnetic drag and thus boost mass growth due to the longer infall time, for sufficiently low $\beta$. Of course, this is sensitive to the relative sequence of stream/CGM formation: if streams are in place before the hot shocked CGM forms, then transverse fields will have little effect.  

   \item We also run stream simulations in a realistic $z\sim2$ halo potential, similar to \citet{Aung2024}. Similar to those authors, we find that streams can remain intact and transport fresh gas to the galaxy at halo centers. However, the detailed behavior of streams (overall mass flux, and infall velocities) depends on their radius, since $t_{\rm grow}/t_{\rm ff} \propto R$. Thus, smaller streams grow faster and experience more accretion braking. For typical stream sizes, the mass flux does not increase significantly. 

   \item Our overall conclusion regarding streams is that while they may be harder to grow than clouds, they are also surprisingly difficult to destroy (in the sense of a declining cold gas mass flux). In the hydrodynamic case, they rapidly accelerate to supersonic velocities, where the Kelvin-Helmholtz instability is quenched, and the destruction time by body modes, which is several times the sound crossing time $\sim 4 R_{\rm s}/c_{\rm s,c}$, is comparable to or longer than the infall time. Magnetic fields further promote survival for the reasons mentioned above. We caution that further exploration of parameter space is warranted, since at high velocities the reduced stream size (for $\rho v R^{2} \approx$ const, $v \propto R^{-1/2}$ for constant density) can reduce the body mode destruction time in comparison to the infall time. However, it is clear that survival criteria can differ significantly from wind-tunnel simulations. 

   \item Finally, we have confirmed the accuracy of the analytic model \citep{tan23-gravity} for hydrodynamic infalling clouds in a realistic potential (equations \ref{eqn:momentum}-- \ref{eqn:dzdt}). We developed a similar analytic model for hydrodynamic infalling streams (equation \ref{eqn:stream_eq1}-\ref{eqn:stream_eq2}), and confirmed they agree well with simulations. We have not developed similar expressions for the MHD case, which would require developing time-dependent expressions for B-field strength in mixing and/or draping layers. However, we have confirmed that if we use simulation results for the draped magnetic field in the transverse field case, analytic expressions for magnetic drag ($F_{\rm drag, MHD} = \rho_{\rm hot} vR^2(1+(v_A/v)^2)$) agree well with the simulation results.

\end{itemize}


Cosmological simulations often show tangled magnetic field configurations for infalling cold streams and clouds. This would naturally result in the final outcome to be a mixture of the aligned and transverse field cases (since these represent the two extremes). In the future we aim to implement a tangled magnetic field configuration to examine their effect on survival and growth. Furthermore, we have neglected thermal conduction and cosmic rays in our studies. It is possible that conduction might have a role in modifying some of the mixing layer properties; however, these have been shown to not be important for the bulk dynamics \citep{tan21}. 
Cosmic rays could also pressurize the mixing layers, thereby changing gas densities and cooling as well as mass growth times. 
We leave this for future work.

\begin{acknowledgments}
We thank Navin Tsung, Zirui Chen and Max Gronke for interesting discussions. The simulations were carried out on the Stampede2 cluster through the ACCESS grant PHY230102. We acknowledge support from NASA grant 19-ATP19-0205 and NSF grant AST240752. NM acknowledges support from from BSF grants 2020302 and 2022281 and NSF-BSF grant 2022736. This research was supported in part by NSF grant PHY-2309135 to the Kavli Institute for Theoretical Physics (KITP). This work was performed in part at the Aspen Center for Physics, which is supported by National Science Foundation grant PHY-2210452.
\end{acknowledgments}

\bibliography{refs_ish,master_references}{}
\bibliographystyle{aasjournal}




\end{document}